\documentclass[aps,preprint,superscriptaddress, 11pt]{revtex4-1}

\usepackage[utf8]{inputenc}
\usepackage{mathtools}

\usepackage{graphicx}
\usepackage{dcolumn}
\usepackage{bm}
\usepackage{makecell}
\usepackage{multirow}
\usepackage{comment}
\usepackage{subfig}
\usepackage[dvipsnames]{xcolor,colortbl}
\usepackage{booktabs} 

\PassOptionsToPackage{hyphens}{url}
\usepackage{hyperref}
\hypersetup{colorlinks=true,linkcolor=blue, linktocpage, urlcolor=blue, citecolor=black}

\begin{document}
\title{Mapping the NFT revolution: \\ market trends, trade networks, and visual features}

\date{\today}

\author{Matthieu Nadini}
\affiliation{Department of Mathematics, City University of London, EC1V 0HB, London, UK}
\affiliation{The Alan Turing Institute, British Library, 96 Euston Road, NW12DB, London, UK}
\author{Laura Alessandretti}
\affiliation{Technical University of Denmark, DK-2800 Kgs. Lyngby, DK}
\author{Flavio Di Giacinto}
\affiliation{Department of Mathematics, City University of London, EC1V 0HB, London, UK}
\affiliation{Department of Neuroscience, Catholic University of the Sacred Heart, Rome, IT}
\author{Mauro Martino}
\affiliation{IBM Research, Cambridge MA, USA}
\author{Luca Maria Aiello}
\affiliation{IT University of Copenhagen, DK}
\author{Andrea Baronchelli}\email{To whom correspondence should be addressed: abaronchelli@turing.ac.uk}
\affiliation{Department of Mathematics, City University of London, EC1V 0HB, London, UK}
\affiliation{The Alan Turing Institute, British Library, 96 Euston Road, NW12DB, London, UK}
\affiliation{UCL Centre for Blockchain Technologies, University College London, London, UK.}

\begin{abstract}
\textbf{Non Fungible Tokens (NFTs) are digital assets that represent objects like art, collectible, and in-game items. They are traded online, often with cryptocurrency, and are generally encoded within smart contracts on a blockchain. Public attention towards NFTs has exploded in 2021, when their market has experienced record sales, but little is known about the overall structure and evolution of its market. Here, we analyse data concerning 6.1 million trades of 4.7 million NFTs between June 23, 2017 and April 27, 2021, obtained primarily from Ethereum and WAX blockchains. First, we characterize statistical properties of the market. Second, we build the network of interactions, show that traders typically specialize on NFTs associated with similar objects and form tight clusters with other traders that exchange the same kind of objects. Third, we cluster objects associated to NFTs according to their visual features and show that collections contain visually homogeneous objects. Finally, we investigate the predictability of NFT sales using simple machine learning algorithms and find that sale history and, secondarily, visual features are good predictors for price. We anticipate that these findings will stimulate further research on NFT production, adoption, and trading in different contexts.}
\end{abstract}

\maketitle

\section{Introduction}

``WTF are NFTs? Why crypto is dominating the art market'' is the title of the February 21, 2021 episode of \textit{The Art Newspaper} podcast~\cite{podcast}, signalling both the impact of Non Fungible Tokens (NFTs) on the art world and the novelty they represent for most of the general public. The revolution is not confined to the art market. While NFT adoption in gaming has already reached a certain maturity, for example concerning the trade of in-game objects, different other industries, especially those involved with the production of digital content such as music or video, are experimenting with the technology. Overall, in the first four months of 2021, the NFT volume has exceeded 2 billion USD, ten times more than the entire NFT trading volume in 2020~\cite{nonfungible_report}.

So, what's an NFT? An NFT is a unit of data stored on a blockchain that certifies a digital asset to be unique and therefore not interchangeable, while offering a unique digital certificate of ownership for the NFT~\cite{evans2019cryptokitties}.
More broadly, an NFT allows to establish the ``provenance'' of the assigned digital object, offering indisputable answers to such questions as who owns, previously owned, and created the NFT, as well as which of the many copies is the original. Several types of digital objects can be associated to an NFT including photos, videos, and audio. NFTs are now being used to commodify digital objects in different contexts, such as art, gaming, and sports collectibles. Originally NFTs were part of the Ethereum blockchain but increasingly more blockchains have implemented their own versions of NFTs~\cite{blockchain_NFT}.

The first popular example of NFTs is CryptoKitties, a collection of artistic images representing virtual cats that are used in a game on Ethereum that allows players to purchase, collect, breed, and sell them on Ethereum~\cite{kitties_main_page}. In December 2017, CryptoKitties congested the Ethereum network~\cite{eth_slowdown}. By many considered a chief example of the irrationality driving the cryptocurrency market in 2017~\cite{kitty_madness}, CryptoKitties remained the only popular example of NFTs for almost two years. In July 2020, the NFT market started to grow~\cite{nonfungible_report} and attracted a huge attention in March 2021, when the artist known as Beeple sold an NFT of his work for \$69.3 million at Christie's~\cite{christie_nft_sold}. The purchase resulted in the third-highest auction price achieved for a living artist, after Jeff Koons and David Hockney~\cite{beeple_third}. Several other record sales followed~\cite{Most_expensive_sales, Most_expensive_sale_cryptopunk}: three Cryptopunks---a collection of $10\,000$ unique automatically generated digital characters---were sold at \$11.8, \$7.6, and \$7.6 million dollars, respectively; the first tweet was sold at \$2.9 million dollars; and the Auction Winner Picks Name, an NFT with music video and dance track, sold at \$1.33 million dollars. The profitability of NFTs has motivated celebrities to create their own NFTs, with collectibles of NBA and famous football players getting sold for hundreds of thousands dollars~\cite{Trading_collectibles_NFT}.

Research on NFTs is still limited, and focuses mostly on technical aspects, such as copyright regulations~\cite{evans2019cryptokitties}; components, protocols, standards, and desired properties~\cite{wang2021non}; new blockchain-based protocols to trace physical goods~\cite{westerkamp2018blockchain}; and the implications that NFTs have on the art world~\cite{whitaker2019art, van2021artist}, in particular as they allow to share secondary sale royalties with the artist. Empirical studies aiming at characterizing properties of the market have focused on a limited number of NFT collections, such as CryptoKitties~\cite{serada2020cryptokitties, sako2021fairness}, Cryptopunks, and Axie~\cite{dowling2021non}, or on a single NFT market, such as Decentraland~\cite{dowling2021fertile,dowling2021non} or SuperRare ~\cite{franceschet2020art, NFT_NYT_Barabasi}. These analyses revealed that the digital abundance of NFTs in digital games has led to a substantial decrease of their value~\cite{serada2020cryptokitties}, and that, even if NFT prices are driven by the prices of cryptocurrencies~\cite{dowling2021non}, the NFT market could be prone to speculation~\cite{sako2021fairness,dowling2021fertile}. Further, it was shown that NFTs valued by experts are more successful~\cite{franceschet2020art}, and that, based on $16\,000$ NFTs sold on the SuperRare market, the structure of the the NFT co-ownership network is highly centralized, and small-world-like~\cite{NFT_NYT_Barabasi,barrat2000properties}. 

In this paper, we provide a first comprehensive quantitative overview of the NFT market. To this end, we analyse a large dataset including 6.1 million trades of 4.7 million NFTs in 160 cryptocurrencies, primarly Ethereum and WAX, and covering the period between June 23, 2017 and April 27, 2021. We start by analysing the overall statistical properties of the NFT market and its evolution over time. Then, we study the network of interactions between NFT traders, and the network of NFT assets. NFTs are further clustered based on their visual features. Finally, we present the results of regression and classification models predicting the occurrence of NFT secondary sales and their price. 

We break down our analysis by NFT categories, which are classified by manual inspection, with references to the classification proposed by NonFungible Corporation~\cite{nonfungible_data}, a specialized company that track NFTs sales, and OpenSea~\cite{OpenSea_assets}, one of the largest NFT marketplace. However, the exact classification of different categories in which NFTs are used is outside of the scope of the present paper. For example, \emph{Art} objects can be in some cases classified as \emph{Collectibles}, while some \emph{Game} objects may present sophisticated aesthetic and cultural properties that may qualify them as \emph{Art}.

\section{Results}

\begin{figure}[h]
\includegraphics[width=\textwidth]{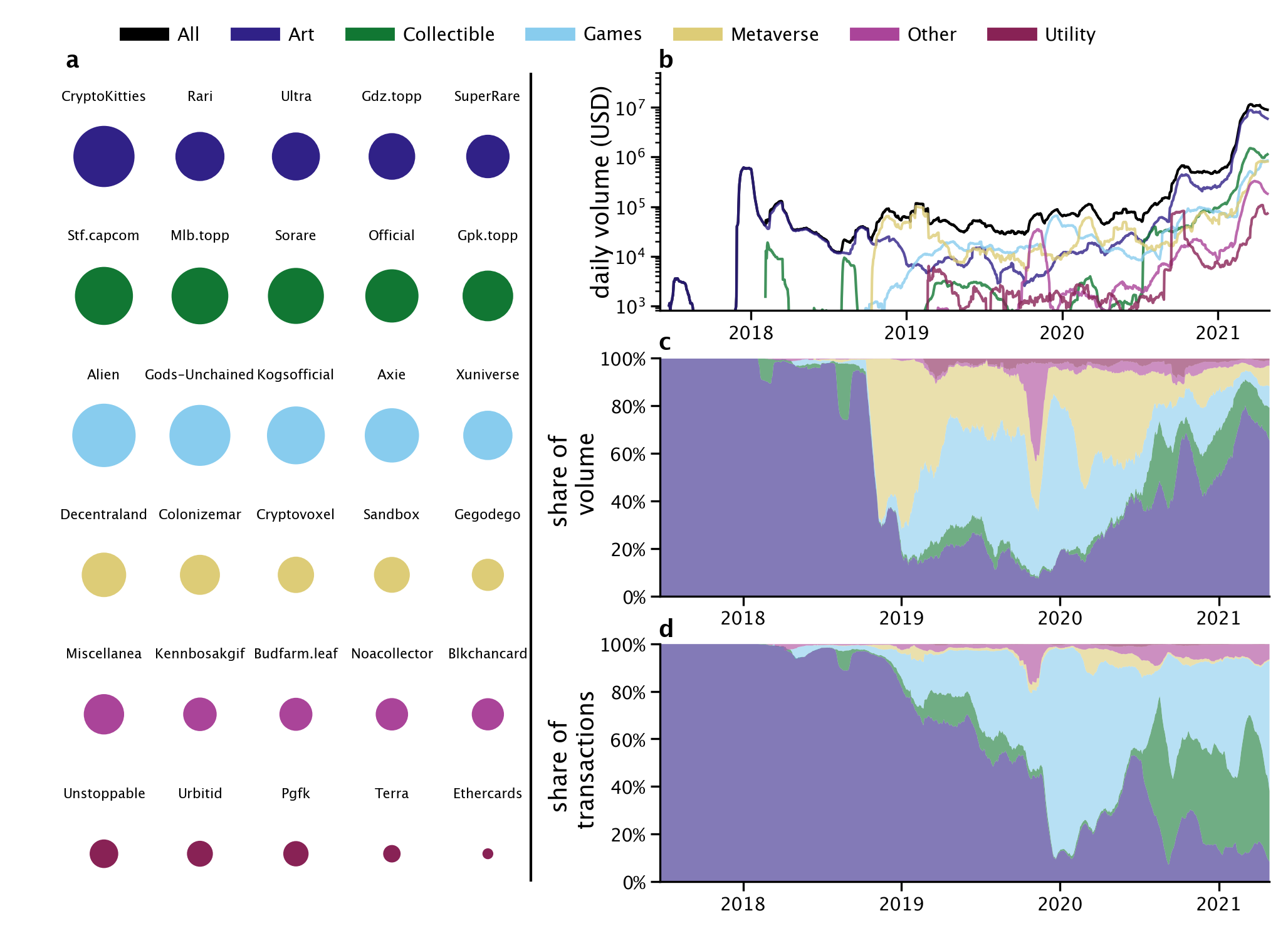}
\caption{\textbf{Description of the NFT landscape.} \textbf{(a) Top 5 NFTs collections (by number of assets) organized by category. The size of each circle is proportional to the number of assets in each collection. \textbf{(b)} Daily volume (in USD) exchanged over time for each category and for all assets (see legend). Days with volume below 1\,000 USD are not shown. \textbf{(c)} Share of volume traded by category. \textbf{(d)} Share of transactions by category. Results in these panels are averaged over a rolling window of 30 days.}}
\label{Fig1}
\end{figure}

\subsection{The NFT market} 
\label{The_NFT_Market}

Items exchanged on the NFT market are organized in \emph{collections}, sets of NFTs that, in most cases, share some common features. Collections can be widely different in nature, from sets of collectible cards, to selections of art masterpieces, to virtual spaces in online games. Most collections can be categorised in six categories: \emph{Art}, \emph{Collectible}, \emph{Games}, \emph{Metaverse}, \emph{Other}, and \emph{Utility} (see also Figure~\ref{FigS1}). 
We show the top $5$ collections in terms of number of unique assets ($n$) for each category (see Figure~\ref{Fig1}a).

Following an initial rapid growth in late 2017, when CryptoKitties collection gained worldwide popularity, the size of the NFT market has remained substantially stable until mid 2020, with an average of $\sim 60\,000$ US dollars traded daily (see Figure~\ref{Fig1}b). Starting from July 2020, the market has experienced a dramatic growth, with the total volume exchanged daily surpassing $\sim10$ million US dollars in March 2021, thus becoming $150$ times larger than it was $8$ months earlier.  

We measured to what extent different NFTs categories contribute to the size of the whole NFT market. Until the end of $2018$, the market was fully dominated by the \emph{Art} category, and in particular by the CryptoKitties collection. 
From January 2019, other categories started gaining popularity, both in terms of total volume exchanged (see Figure~\ref{Fig1}b-c) and number of transactions (see Figure~\ref{Fig1}d).
Overall, in the period between January 2019 and July 2020, $\sim 90\%$ of the total volume exchanged on NFT was shared by the \emph{Art}, \emph{Games}, and \emph{Metaverse} categories, contributing $18\%$, $33\%$, and $39\%$ respectively. Starting from mid July 2020, the market volume has been largely dominated by NFTs categorized as \emph{Art}, which, since then, have contributed $\sim 71\%$ of the total transaction volume, followed by \emph{Collectible} assets accounting for $12\%$. Importantly, however, the market composition is quite different when considering the number of transactions. Since July 2020, the most exchanged NFTs belong to the categories \emph{Games} and \emph{Collectible}, which account for $44\%$ and $38\%$ of transactions. Instead, only $10\%$ of transactions are related to NFTs categorized as \emph{Art}. Overall, we observe that the share of volume spent in \emph{Art} has been growing since $2020$, while its share of transactions has been decreasing (Figure~\ref{Fig1}d). The discrepancy between volume and transactions reveals that prices of items categorized as \emph{Art} are higher, on average, compared to other categories. 

\begin{figure}[h]
\includegraphics[width=\textwidth]{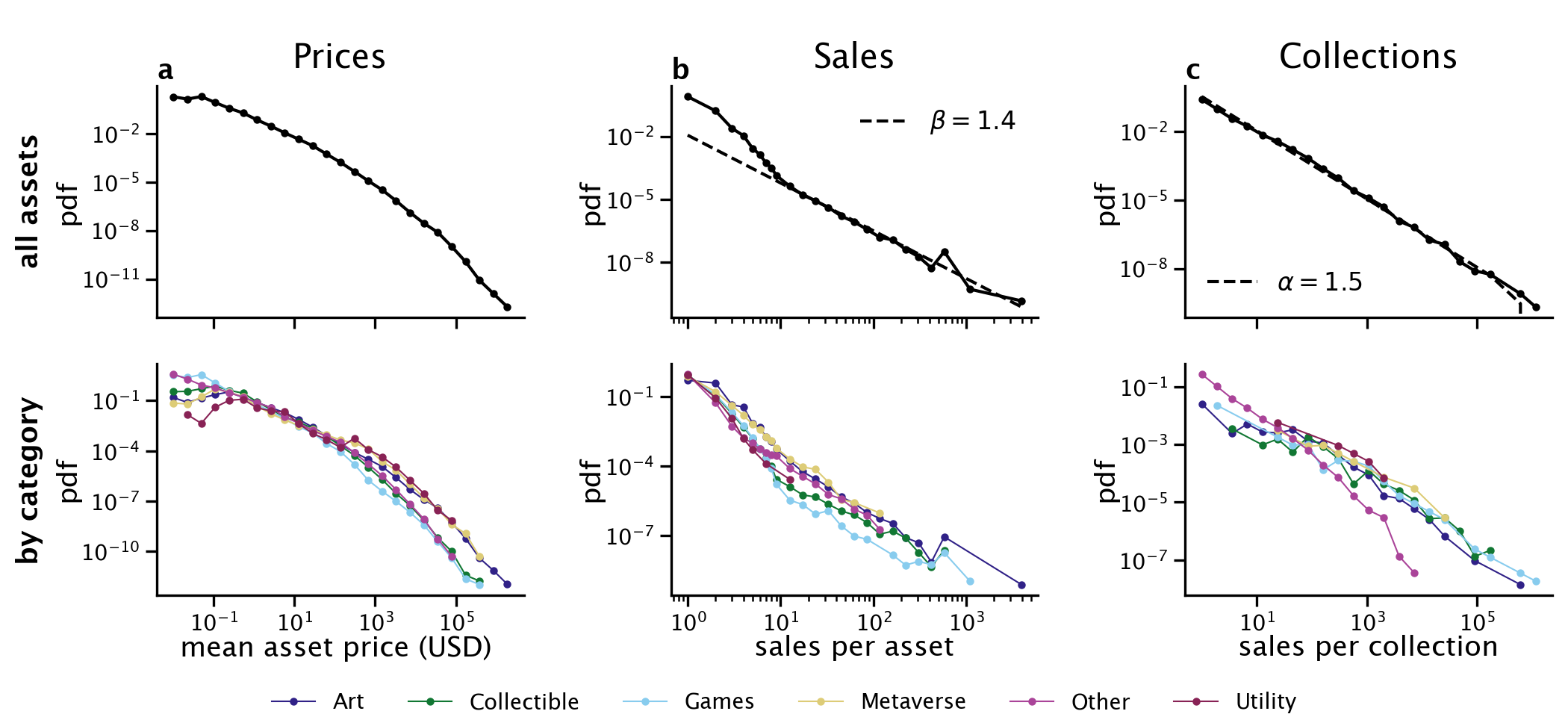}
\caption{\textbf{Statistical properties of the NFT market.} \textbf{(a)} Distribution of the average price (USD) for all NFTs (top) and by NFT category (bottom).  \textbf{(b)} Distribution of number of sales per NFT for all NFTs (top) and by category (bottom). The dashed line is a power law fit $P(s) \sim s^{\beta}$, with $\beta=-1.4$, where $s$ is the number of sales. \textbf{(c)} Distribution of number of assets per collection for all NFTs (top) and by  category (bottom). The dashed line is a power law fit $P(n) \sim n^{\alpha}$, with $\alpha=-1.5$, where $n$ is the number of unique assets.}
\label{Fig3}
\end{figure}

We dig further into these differences by looking at the distribution of NFT prices across categories (see Figure~\ref{Fig3}a), which we find to be broadly distributed. We observe that the average sale price of NFTs is lower than $15$ dollars for $75\%$ of the assets, and larger than $1\,594$ dollars, for $1\%$ of the assets. Considering individual categories, NFTs categorized as \emph{Art}, \emph{Metaverse}, and \emph{Utility} reached higher prices compared to other categories, with the top $1\%$ of assets having average sale price higher than $6\,290$, $9\,485$, and $12\,756$ dollars respectively. Note that these categories are different in sizes, so $1\%$ of assets corresponds to $8\,593$, $472$, and $78$ NFTs in the \emph{Art}, \emph{Metaverse}, and \emph{Utility} categories, respectively. The highest prices so far were reached by assets categorized as \emph{Art}, with $4$ NFTs that were sold for more than $1$ million dollars.

To assess the market activity, we measured how often individual assets are traded. Here, we refer to the first time an asset is sold as the asset's \emph{primary sale}, and to all other sales as \emph{secondary sales}. All assets considered in this study had a primary sale, but only $\sim 20\%$ of them had a secondary sale (see Figure~\ref{fig:fraction_nft_with_resale}). We observe that the tail of the distribution of number of sales $s$ per asset, for $s \geq 10$, is well characterized by a power-law function $P(s) \sim s^{\beta}$, with $\beta=-1.4$, estimated following~\cite{clauset2009power} (see Figure~\ref{Fig3}b). When looking at different categories, the distribution of number of sales is affected by cut-off values. For example, the maximum number of sales for assets in the \emph{Utility} category is 12, while an asset in the \emph{Games} category is sold more than a thousand times, and an asset in the \emph{Art} category more than five thousands times. Note that only $0.07\%$ of all assets are sold more than $10$ times. Also, the size of collections $n$ is well described by a power-law function $P(n) \sim n^{\alpha}$, with $\alpha=-1.5$ (see Figure~\ref{Fig3}c), implying the distribution of sizes is broad. We find that $\sim75\%$ of collections comprise less than $37$ unique assets, and $\sim1\%$ have more than $10\,400$ unique assets.

\begin{figure}[h]
\includegraphics[width=\textwidth]{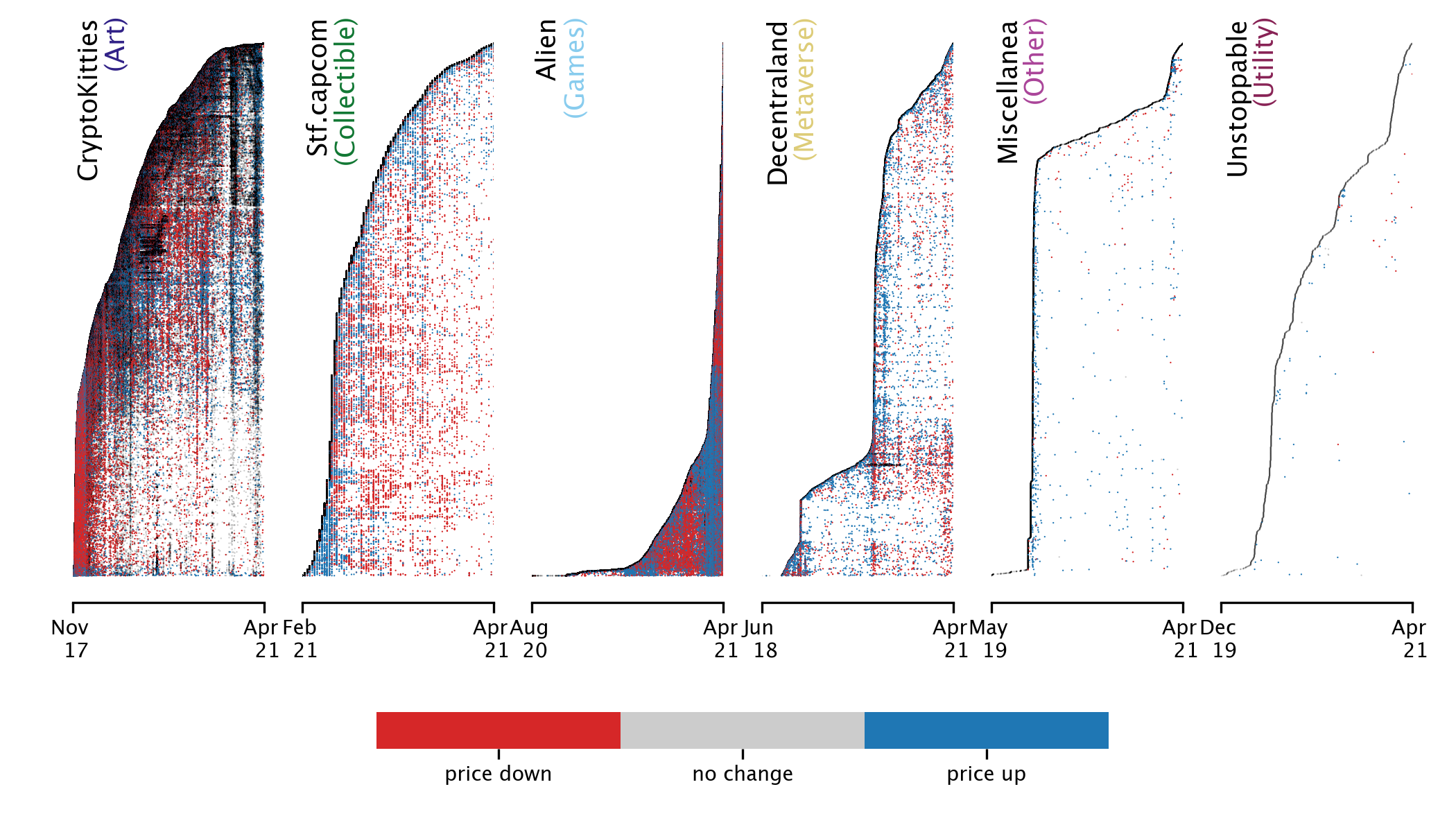}
\caption{\textbf{Secondary sale prices.} Sales over time for the top collection in terms of number of sales in each NFT category (CryptoKitties, Stf.capcorn, Alien, Decentraland, Miscellanea, and Unstoppable). Each horizontal line represents an NFT and each dot a sale. Sales are coloured based on the change in price compared to previous sale (see colourbar).}
\label{Secondary_sale_prices}
\end{figure}

Temporal patterns of secondary sales are unique for each collection, as evidenced by considering the top collection in each category (see Figure~\ref{Secondary_sale_prices}). For example, when Cryptokitties emerged in 2017, secondary sale prices were typically lower than the price of their first sale. More recently in 2021, instead, their secondary sale prices have gone up because of an increase in the number of potential customers. Other collections, like Alien, alternated periods when secondary sale prices went down and period when they went up. In the Unstoppable collection secondary sales are rare because NFTs correspond to web domains secured by blockchain technology. In 2017, secondary sale price were lower than primary in 66\% of the cases, while in 2021 only 27\% of secondary sales had lower prices than the primary one.

\subsection{The networks of NFT trades} 
\label{The_networks_of_NFT_trades}

How do traders interact with each other? Are there central actors? We approach these questions adopting a network science approach~\cite{barrat2000properties,barrat2004architecture}.
We consider the \emph{network of trades}, where nodes are traders, a directed link from a trader to another exists if the former (the buyer) purchases at least one NFT from the latter (the seller). Each link has a weight corresponding to the total number of items that the buyer bought from the seller.

\begin{figure}[h]
\includegraphics[width=\textwidth]{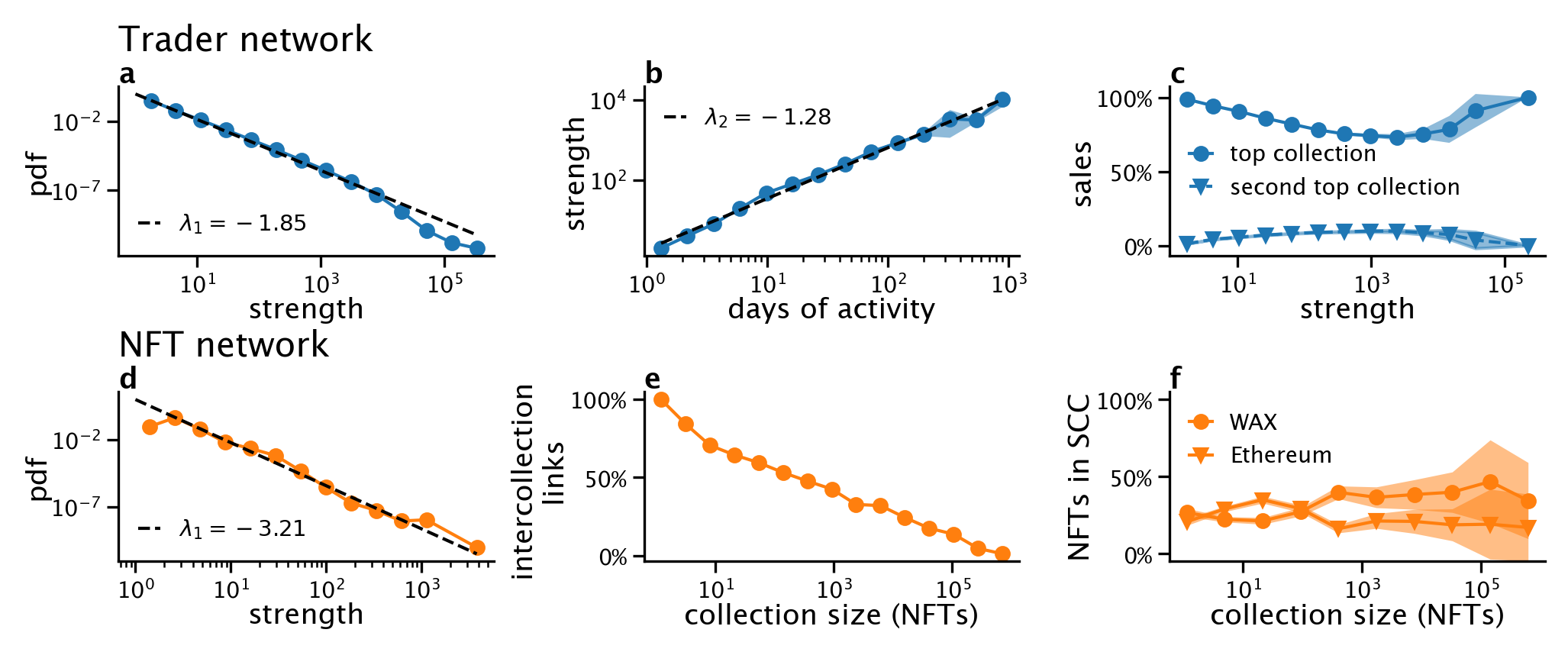}
\caption{\textbf{Key network properties.} \textbf{(a)} Pdf of the traders' strength. \textbf{(b)} Traders' strength as a function of the number of days of activity. \textbf{(c)} Percentage of transaction traders make toward their top and second-top NFT collections. \textbf{(d)} Pdf of the NFTs' strength. \textbf{(e)} Percentage of transactions between NFTs in different collections as a function of the size of the collection. \textbf{(f)} Percentage of NFTs belonging to the first and second largest strong connected component (SCC). Solid curves in panels \textbf{(b)-(c)-(e)-(f)} represent average values, while respective bands the 95\% confidence interval.} 
\label{Fig2_network}
\end{figure}

\begin{figure}[h]
\includegraphics[width=\textwidth]{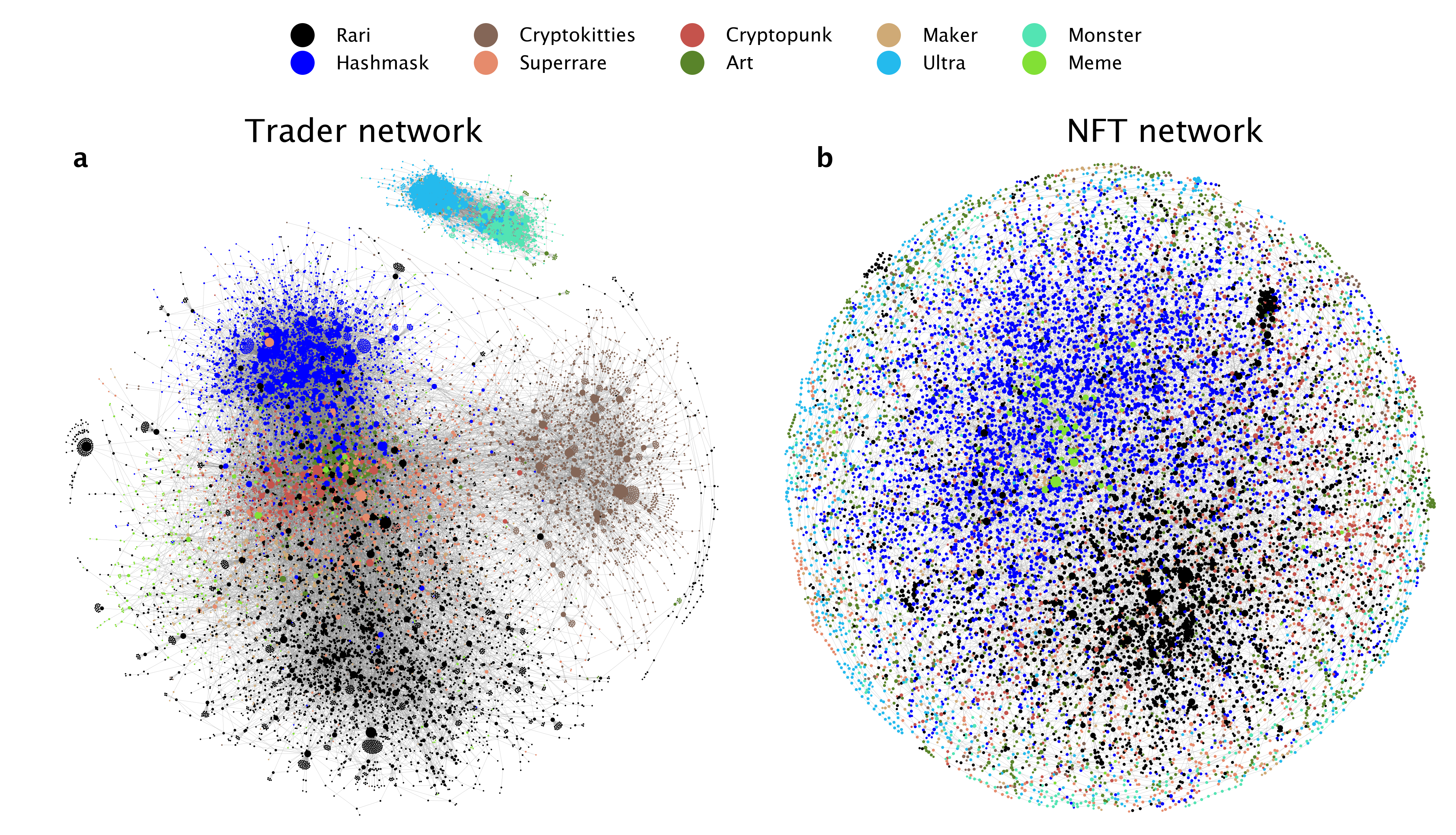}
\caption{\textbf{Networks visualization.} \textbf{(a)} Trader network, where nodes represent traders and links sales between a pair of them. \textbf{(b)} NFT network, where nodes represent NFTs and links when a pair of NFTs is purchased in ``sequence''. For visualization purposes, we selected the 10 top collections in the \emph{Art} category on February 2021. Visualization is done using Netwulf~\cite{aslak2019netwulf}.}
\label{FigNetwork_visualization}
\end{figure}

First, we study the behaviour of individual NFT traders by focusing on properties of the nodes. We find that traders activity is highly heterogeneous: the strength of traders (nodes) $s$, defined as the total number of purchases and sales made by each trader, is distributed as a power law $P(s)\sim s^{\lambda_1}$ with exponent $\lambda_1 = -1.85$ (Figure~\ref{Fig2_network}a), such that the top 10\% of traders alone perform 85\% of all transactions and trade at least once 97\% of all assets.  Further, we find a superlinear relation between the strength of a trader and the total number of days of activity $d$, with $s\sim d^{\lambda_2}$ and $\lambda_2 = 1.28$. This result reveals that the average number of daily trades is larger for traders active over long periods of time. Traders are also specialized: measuring how individuals distribute their trades across collections, we find that traders perform at least 73\% of their transactions in their top collection, while at least 82\% in their top two collections combined. The relation between strength and specialization is not monotonic: the most specialized traders have either few (less than ten) or many (more than ten thousands) transactions (see Figure~\ref{Fig2_network}c). A specialized trader is the one with Ethereum address ``0xfc624f8f58db41bdb95aedee1de3c1cf047105f1'', that exchanges tens of thousands of CryptoKitties. Similar relationships hold when buying and selling behaviours are considered separately (see Figure~\ref{FigS2_network}). \\
Secondly, we turn to properties of the network links, describing interactions between pairs of traders. We find that the distribution of link weights is well characterized by a power law distribution, with the top 10\% of buyer-seller pairs contributing to the total number of transactions as much as the remaining 90\% (see Figure~\ref{FigS2_network}a). An interesting question is whether traders connect preferentially to traders that have similar strength. We tackle this question by studying the assortativity coefficient $r$~\cite{newman2003mixing}, that measures the correlation between the sum of the weights of all outgoing links (the outgoing strength) of a given node with the average sum of the weights of incoming links (the incoming strength) of its neighbours. We find that the assortativity, which takes value $r=-0.024$, is close to the null value zero, implying that traders do not connect to other traders based on the similarity of their connection patterns.\\
Finally, we focus on the network structure. Building upon the result that traders are specialized, we assign each trader to their top collection, and we study the modularity~\cite{clauset2004finding} of the network under this partition of nodes. The modularity is a metric bounded between $-0.5$ and $1$, which is positive when the density of links among nodes assigned to the same partition is larger than it would be expected by chance. We find that the modularity $Q$ of the collections partition is $Q = 0.613$, significantly higher than what expected from a random network $Q = 0.0823 \pm 0.0001$ (see Section \ref{sec:random_network}). It reveals that the collections well represent the underlining network structure, where traders specialized in a collection tends to buy and sell NFTs with other traders specialized in the same collection. A visual representation of the trader network including the \emph{Art} category on February 2021 shows the clusters formed by NFT traders specialized in the same collection (see Figure~\ref{FigNetwork_visualization}a).

We now turn to the exploration of how NFTs are connected to one another. To this end, we construct the \emph{network of NFTs}, where nodes are NFTs and a directed link exists between two NFTs that are purchased ``in sequence'', e.g. a link is created from an NFT to another when a buyer purchases the former and then the latter, with no purchases between the two (see Section~\ref{Trader_NFT_networks} for more details). Rather than linking all NFTs ever traded by the same trader, this choice allows to understand the relations between NFT that are semantically similar, because they are bought by the same trader in approximately the same period of time. Further, it ensures that the network structure is not dominated by large cliques.

The distribution of NFTs strength decays as a power law with exponent $\lambda_3 = -3.21$ (see Figure~\ref{Fig2_network}d). Note that the strength of NFTs is different to the total number of sales per NFT (previously shown in Figure~\ref{Fig3}b), due to how the network is constructed. In fact, when two NFTs are purchased simultaneously, this creates two links for each of the two nodes (one ingoing and one outgoing). 
The next question we ask is: which NFTs are connected to one another? We find that NFTs in small collections tend to be bought in sequence with NFTs in other collections (see Figure~\ref{Fig2_network}e). On the contrary, NFTs in large collections, like CryptoKitties or Gods-Unchained, tend to be bought in sequence with NFTs in the same collection. \\
What are the implications of this behaviour on the NFT network structure?
We investigate the relation between the structure of the NFT network and NFTs collections, by studying the modularity~\cite{clauset2004finding} of the network under the partition of NFTs (nodes) into NFT collections. We find that the modularity $Q$ of the collections partition is $Q = 0.80$, significantly higher than what expected
from a random network $Q = 0.1110 \pm 0.0001$. It reveals that (i) the network is clustered and (ii) the collections well represent the underlining community structure. By further exploring the relationship between traders' behaviour and NFT networks structure, we unveil that, while the NFT network is clustered, communities are not isolated. That is, some traders buy or sell assets belonging to multiple collections. The network of NFTs has two strongly connected components (SCC)~\cite{nuutila1994finding}, defined as groups of nodes such that, starting from a given NFTs, it is possible to reach any other NFTs in the SCC following directed links. The largest SCC include NFTs traded in the WAX blockchain, consisting of 35\% of all NFTs, while the second largest includes NFTs traded in the Ethereum blockchain, consisting of 20\% of all NFTs. While the high network modularity reveals that traders tend to purchase assets from the same collection in sequence, the presence of very large SCCs reveals that there are less frequent sequences of purchases in different collections. A visual representation of the NFT network including the \emph{Art} category on February 2021 exemplifies our findings, where NFTs, albeit surrounded by other NFTs in the same collection, tend to form a sparser structure (see Figure~\ref{FigNetwork_visualization}b).

We then study the networks consisting of assets in the same category and blockchain (see Figure~\ref{Fig_networks_categories} and~\ref{Fig_networks_blockchain}). We find that key results presented above, including the shape of the strength distributions, hold across categories (see Figure~\ref{Fig_networks_categories}a-b-d-e and~\ref{Fig_networks_blockchain}a-b-d-e). Also in this case, we find that traders, independently from the category considered, are specialized: the fraction of individual trades in the top collection is included between 59\%, for the \emph{Other} category, and 98\%, for the \emph{Utility} category (see Figure~\ref{Fig_networks_categories}c). Similarly, the fraction of individual trades in the top collection is 70\% for the WAX blockchain and 91\% for the Ethereum blockchain category (see Figure~\ref{Fig_networks_blockchain}c). Relative to the number of total NFTs in each category, the WAX component contains 55.0\% of all NFTs labeled as \emph{Collectible}, but only the 0.06\% of all NFTs labeled as \emph{Utility}. On the contrary, the Ethereum component has the 54.8\% of all \emph{Art}, but only the 10.6\% of \emph{Games}.

\subsection{Visual features} 
\label{Graphical_features}

NFTs are linked to digital assets of different types, including videos, text, animated GIFs, and audio. Currently, the most popular NFTs are images~\cite{Most_expensive_sales, Most_expensive_sale_cryptopunk}. We select NFTs associated with images and take a snapshot of animated GIFs, and analyse them with the pre-trained convolution neural network AlexNet. AlexNet extracts from an image a vector of 4\,096 values that is a dense representation of the image's visual features. With this representation, vectors extracted from images that are visually similar are close in the vector space. To quantify the visual difference between pairs of pictures, we calculated the cosine distance (CD) between them, a value that goes from zero (for identical images) to one (for highly different images). We measured such distance between pictures within the same collection and across collections. 

\begin{figure}[h]
\includegraphics[width=\textwidth]{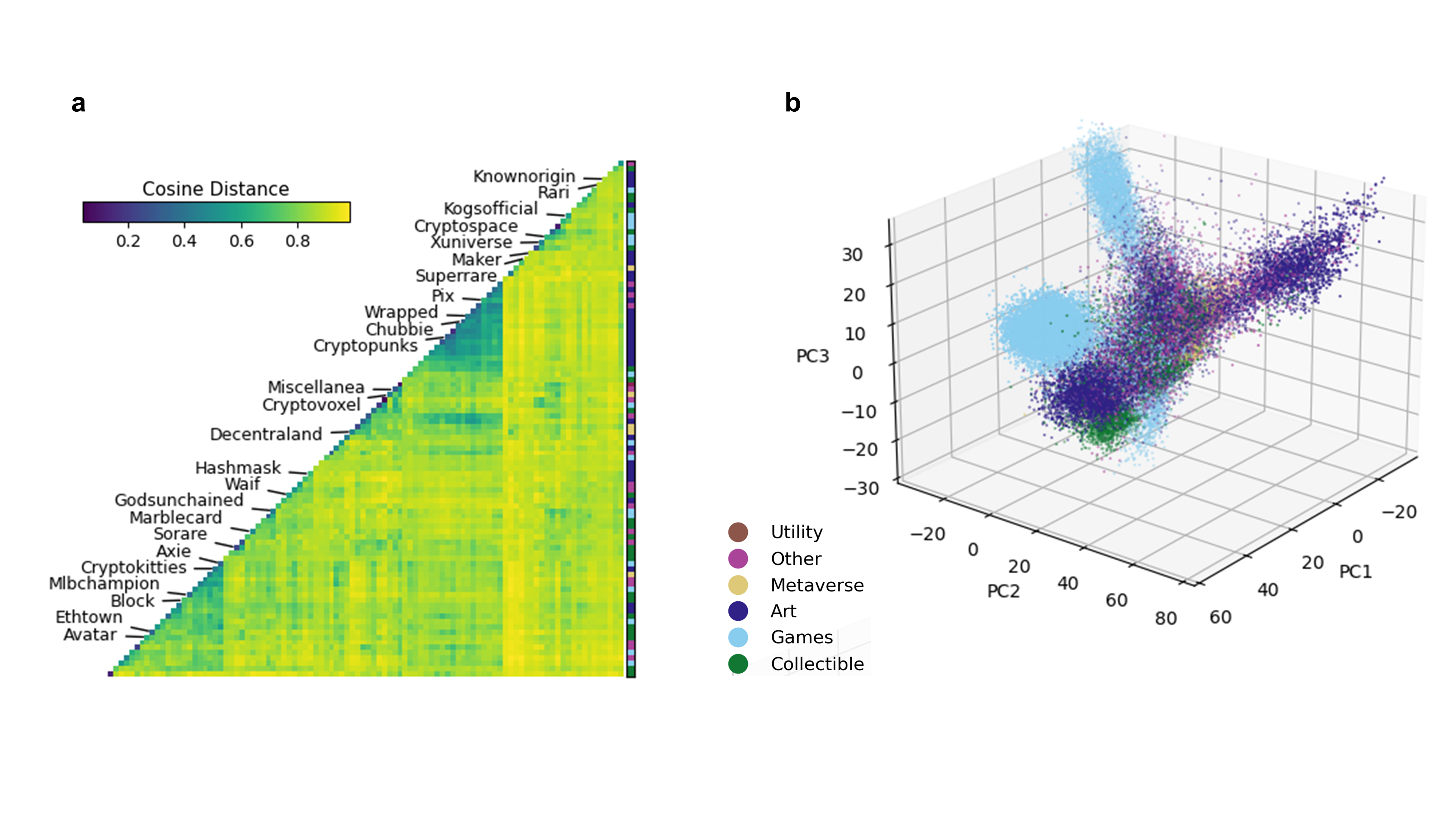}
\caption{\textbf{Visual features representation.} \textbf{(a)} Cosine distance of graphical digital objects between items grouped by collections and categories (coloured bands on the right), recognising aesthetical similarities and uniformity between and within these groups.
For visualization purposes, we selected the largest 98 collections in our dataset. \textbf{(b)} The dimensionality reduction of AlexNet vectors by PCA and their visualization in the PC1, PC2 and PC3 space, broken down by NFT categories, demonstrate the presence of graphically uniform clusters. For visualization purposes, we downsampled the digital objects associated with the CryptoKitties and Sorare collections, which alone constitute the 61\% of the whole dataset.} 
\label{PCAandClusters}
\end{figure}

The average CD calculated between items which belong to the same collection is significantly lower ($\mu = 0.59$, $\sigma = 0.20$) compared to the one obtained for objects from two different collections ($\mu = 0.87$, $\sigma = 0.06$), confirming an intra-collection graphical homogeneity. Figure~\ref{PCAandClusters}a shows the matrix of average CD values between all pairs of collections. Values on the diagonal represent the intra-collection CD values, and reveal that most collections have a high degree of homogeneity (e.g., Sorare (CD $ = 0.24$) or Cryptopunks (CD $ =0.33 $)) but some are more heterogeneous (Rarible (CD $ = 0.89$)). In short, many collections have their own style, graphical hallmarks that distinguish them from others. There are also sub-groups of collections, usually within the same category (coloured band in Figure~\ref{PCAandClusters}a), which share some common visual features. This is the case for collections containing pieces of pixel-art, including Chubbie, Cryptopunks and Wrapped Punks, or the similarities observed between Cryptokitties and Axie (Figure ~\ref{PCAandClusters}a). 

To map the images into a lower-dimensional feature space that can be used in practice for prediction and visualization, we apply Principal Component Analysis (PCA) to the AlexNet vectors. PCA uses linear combinations of the 4\,096-dimensional vectors to project them into vectors with an arbitrarily lower number of dimensions and such that the variance of datapoints in the projected space is maximized. Considering the whole sample, which consists of about 1.25 million graphical objects, the first five principal components explain together about the 38.3\% of the total variance, progressively distributed from PC1 to PC5 as follow: 20.3\%, 7.3\%, 4.0\%, 3.8\% and 2.7\%. The PC1 to PC5 scores are used to test the capacity of visual features for predicting sales (see Section \ref{sec:prediction_results}), while PC1, PC2, and PC3 for visually representing the data through a 3D scatter plot and showing intra-categories homogeneity (see Figure~\ref{PCAandClusters}b). 
This can be quantified by looking at the average Euclidean distance in the PC1, PC2, PC3 space between objects of the same category and comparing it to the one calculated among objects of different categories. Considering the whole sample and calculating the distance between all the points, the average value obtained between elements of different categories is 1.67 bigger than for elements of the same category. However, as we already described for the cosine distance in the AlexNet vector space, this is mainly due to the intra-collections homogeneity, as demonstrated calculating the average inter-collection distance which results more than three times (3.17) bigger than the intra-collection distance and secondarily to the presence of intra-categories clusters of similar looking collections. This is most likely caused by the market responsiveness to the success of a collection, which induces other creators to follow the trend and offer variations on the theme. 

\clearpage

\subsection{Predicting sales} 
\label{sec:prediction_results}

To identify the factors associated with an NFT's market value, we fit a linear regression model to estimate the price of primary and secondary sales from different sets of features, calculated considering only the data preceding the day of the NFT's primary sale. The features (whose detailed formulations are provided in Section~\ref{sec:features}) include the degree and PageRank centrality of the buyer and seller in the networks of NFT trades ($k_{buyer|seller}$, $PR_{buyer|seller}$), the principal components of visual features of the object linked to the NFT ($vis_{PCA_{1 \ldots 5}}$), a prior probability of sale within the collection ($p_{resale}$), and the past median price of primary and secondary sales within the collection ($median\:price$).

NFT's price correlates strongly with the price of NFTs previously sold within the same collection (see Figure~\ref{fig:regression_vs_window_before}a). The median sale price of NFTs in the collection predicts more than half of the variance of price of future primary and secondary sales. The prediction is more accurate when the median of the past sale price is calculated over a recent time window preceding the primary sale, e.g., the prior time window of one week is better than considering the entire time window preceding the NFT's primary sale. Similar results, albeit with generally lower correlations, are found when the secondary sale price is the object of the regression (see Figure~\ref{fig:regression_vs_window_before}b). As one would expect, the price of secondary sales is strongly correlated with the price of primary sale, and the predictive power of the variables declines as one attempts to cast a prediction over longer periods of time: $R_{adj}^2=0.90$ when predicting the median secondary sale price over the next week, and falls to $R_{adj}^2=0.77$ when extending the prediction over the next 2 years (see Figure~\ref{fig:regression_vs_window_before}c). A similar relation is found between the secondary sale price and the median price of the NFTs collection (see Figure~\ref{fig:regression_vs_window_before}d).

\begin{figure}[h]
\includegraphics[width=\textwidth]{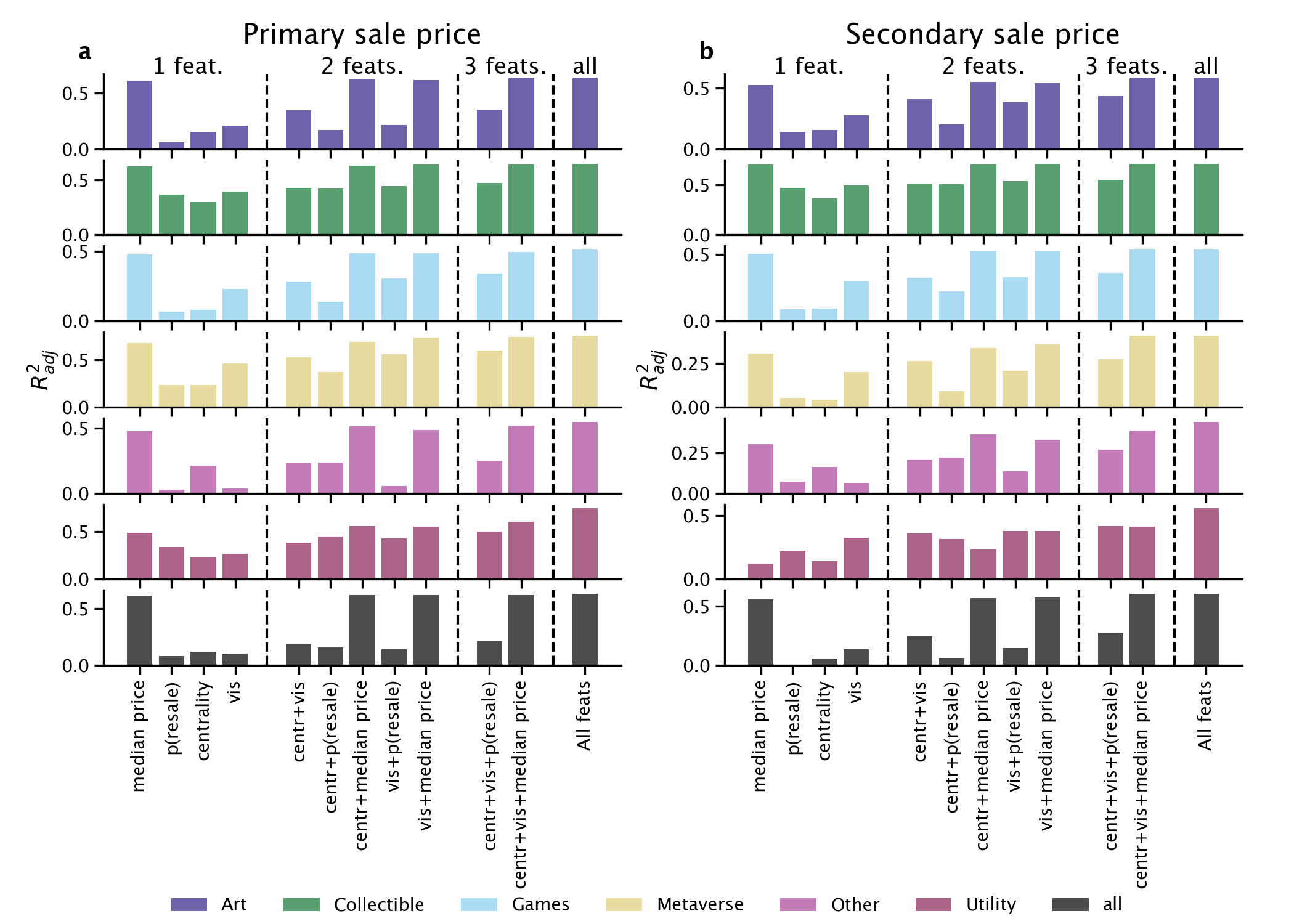}
\caption{\textbf{Regression results.} $R_{adj}^2$ of a linear regression fit to predict the primary price of sale \textbf{(a)} and the median price of secondary sale 1 month after the primary sale \textbf{(b)} from different sets of features. Results are broken down by NFT categories. The abbreviation ``feat(s).'' stands for ``feature(s)''.}
\label{fig:regression_features_breakdown}
\end{figure}

Other features than prior sale history are predictive of future sale and secondary sale prices (see Figure~\ref{fig:regression_features_breakdown}). Centrality measures of the buyer and seller in the trader network ($R_{adj}^2\in[0.05,0.12]$) and visual features of the object linked to the NFT ($R_{adj}^2 \in [0,0.08]$) explain roughly one-fifth to one-fourth of the variance when used in combination ($R_{adj}^2 \in [0.18,0.25]$). When considered in combination with the median price of previous sales, they increase the predictive power by almost 10\% for the secondary sale price ($R_{adj}^2$ from $0.55$ to $0.6$). 
When fitting separate regressions for each category, it becomes apparent that the predictability of future prices and the predictive power of different sets of features varies depending on the NFT category. The \emph{collectible} category is the easiest to predict, with centrality and visual features yielding $R_{adj}^2 \in [0.30,0.36]$ and $R_{adj}^2 \in [0.40,0.50]$, respectively. These two families of features have the largest compound effect in the \emph{Art} category; in the secondary sale price prediction, centrality features boost the predictive power of visual features by more than 50\%. Regression coefficients of individual features for the task of secondary sale price prediction one month after the primary sale are presented in Table~\ref{tab:regression}.

\begin{table}[!t]
\footnotesize
\setlength{\tabcolsep}{5pt}
\begin{center}
\begin{tabular}{lccccccc}
& \multicolumn{7}{c}{\textbf{$\beta$} coefficients}\\
\textbf{Feature}	& \textbf{All} &	\textbf{Art} & \textbf{Collectible}  & \textbf{Games} &  \textbf{Metaverse} & \textbf{Utility} & \textbf{Other}  \\
\hline
\hline
const.         & -0.029 & 0.030  & -0.086 & -0.181 & 0.210 & 2.054 & 0.149 \\
$k_{buyer}$    & -0.018 & 0.022  & -0.032 & -0.132 & -0.078 & -0.010$^{\bullet}$ & -0.207 \\
$k_{seller}$   & -0.166 & -0.211 & 0.000  & 0.026  & 0.166 & 0.198$^{\bullet}$ & -0.347 \\
$PR_{buyer}$   & 0.129  & 0.077  & 0.162  & 0.317  & 0.206 & -0.241$^{\bullet}$ & 0.336 \\
$PR_{seller}$  & 0.302  & 0.367  & -0.031 & -0.066 & 0.009$^{\bullet}$ & -0.382 & 0.459 \\
$p_{resale}$   & 0.029  & -0.041 & 0.079  & 0.023  & 0.046$^{\bullet}$ & 0.465 & 0.251$^{\bullet}$ \\
$median price$ & 0.769  & 0.711  & 0.970  & 0.815  & 0.436 & 0.478 & 0.687 \\
$vis_{PCA_1}$  & 0.098  & 0.153  & 0.049  & 0.174  & 0.175 & -1.136 & 0.021 \\
$vis_{PCA_2}$  & -0.120 & -0.130 & -0.044 & -0.064 & -0.669 & -0.817 & -0.181 \\
$vis_{PCA_3}$  & 0.019  & 0.027  & 0.063  & 0.203  & 0.112$^{\bullet}$ & -1.292 & -0.037$^{\bullet}$ \\
$vis_{PCA_4}$  & 0.040  & 0.028  & -0.003$^{\bullet}$ & 0.130 & -0.018$^{\bullet}$ & -0.911 & -0.116 \\
$vis_{PCA_5}$  & 0.063  & 0.018  & 0.276  & 0.102  & 0.296 & 0.071$^{\bullet}$ & 0.301 \\
\hline
\#NFTs & 407,549 & 251,369 & 69,015 & 78,848 & 2,693 & 314 & 5,297 \\ 
\#Collections & 3307 & 114 & 73 & 48 & 12 & 6 & 3054 \\ 
$R_{adj}^2$ & \textbf{0.6} & \textbf{0.589} & \textbf{0.709} & \textbf{0.535} & \textbf{0.408} & \textbf{0.562} & \textbf{0.44} \\
\end{tabular}
\end{center}
\caption{\textbf{Secondary sale price prediction.} Linear regressions to predict the NFTs' median secondary sale price one month after their primary sale from three families of features: centrality on the trader network ($k$, $PR$), history of sales in the NFT's collection (namely prior probability of secondary sale $p_{resale}$ and median sale price 1 week before the sale $median price$), and visual features ($vis_{PCA_i}$). Regression models were fit to different categories of NFTs independently. For each category, the number of NFTs and collections it contains is reported. The $R_{adj}^2$ is a measure of goodness of fit, and it quantifies the proportion of the data variance explained by the model. The p-values of all $\beta$ coefficients are $<0.01$ except for those marked with $^{\bullet}$, which are all $>0.05$.}
\label{tab:regression}
\end{table}

When predicting secondary sale prices, we consider only those NFTs that were sold in a secondary sale. These NFTs are the minority: less than 10\% are sold at least once within one week after the primary sale, and only about 22\% within one year (see Figure~\ref{fig:fraction_nft_with_resale}). Using the same set of features that we selected for the price regression, we trained AdaBoost~\cite{freund1999short}, a binary classifier, to assess to what extent it is possible to predict whether an NFT will be sold after its primary sale (for more details see in Section~\ref{sec:classification}). We find that this is possible to a certain extent. The prediction is most accurate when training and testing the classifier on \emph{Art} NFTs only ($F1>0.8$), whereas the prediction is less reliable for the other categories ($F1 \in [0.14,0.33]$, see Figure \ref{fig:prediction_features_breakdown}). The median price of the collection is among the strongest predictors, but not always the strongest. The prior probability of sale in the collection is also a strong signal, and centrality and visual features combined can sometimes outperform other feature combinations (e.g., in the \emph{Metaverse} category). Last, the prediction is most accurate when trying to predict the occurrence of a secondary sale over longer periods of time (see Figure~\ref{fig:prediction_allfeats_vs_time}).

\section{Conclusion}
\label{sec:conclusion}

The NFT market is less than four years old and has boomed in 2021. This paper presented the first overview of some key aspects of it by looking at the market history of 6.1 million NFT trades across six main NFT categories including art, games and collectibles. In brief, (i) we analyzed the main properties of the market, (ii) we built and studied the traders and NFTs networks and found that most traders are specialised, (iii) we showed that NFT collections tend to be visually homogeneous, and (iv) we explored the predictability of NFT prices revealing that, while past history is as expected the best predictor, also NFT specific properties, such as the visual features of the associated digital object, help increase predictability.

It is important to highlight the main limitations of our study, which represent also directions for future work. First, we gathered data from a variety of online NFT marketplaces and not directly from the Ethereum or WAX blockchains, so that we have likely missed a number of ``independent'' NFT producers. Second, we mostly adopted an accepted categorisation for the NFTs, which includes a number of arbitrary decisions and could however be further refined (as every categorization). Third, since our primary goal was to provide a general overview of the market, we did not extensively explore all the available methods e.g.,  for the features extraction from images \cite{khan2020survey} and their clustering in a lower-dimensional space \cite{xu2019review}, machine learning for price prediction \cite{alessandretti2018anticipating}, or market modelling \cite{elbahrawy2017evolutionary}. We also did not consider collective attention as measured e.g. from social media or Wikipedia, which can be a further source of information about market behaviour \cite{preis2013quantifying,moat2013quantifying,elbahrawy2019wikipedia}. Fourth, we considered mostly the Ethereum and WAX blockchains, but several other platforms offer smart contracts and NFTs. Finally, our price prediction exercise did not include information about the creator of the (digital) object associated to the NFTs. While this is due mainly to the dataset, and in many cases the identity of the creator is not available or does not exist (e.g., for AI generated images), it is likely that in certain contexts, and specifically for art, this can be an important aspect to consider. 

Overall, NFTs are a new tool that satisfies some of the needs of creators, users, and collectors of a large class of digital and non-digital objects. As such, they are probably here to stay or, at least, they represent a first step towards new tools to deal with property and provenance of such assets. We anticipate that our study will help accelerate new research on NFT in a broad array of disciplines, including economics, law, cultural evolution, art history, computational social science, and computer science. The results will also help practitioners make sense of a rapidly evolving landscape and inform the design of more efficient marketplaces as well as the associated regulation.

\clearpage

\subsection{Data and methods}

We summarize our data collection below and provide a detailed description of our data manipulations in Section~\ref{Data_methods_additional}. 

\subsubsection{Sales data collection}
Our dataset includes only transactions representing purchases of NFTs, whose ownership change following that transaction. We exclude from our analysis any transactions representing the minting of NFTs or  bids during an auction. We track different cryptocurrencies. Etherum blockchain data for the collections SuperRare, Makersplace, Knownorigin, Cryptopunks, and Asyncart were shared by NonFungible Corporation~\cite{nonfungible_data}, a company that tracks historical NFT sales data to build NFT valuations. Other Ethereum blockchain data were downloaded from four open-source APIs: CryptoKitties sales~\cite{Cryptokitties_sales}, Gods-Unchained~\cite{Unchained_Marketplace}, Decentraland~\cite{Decentraland_Marketplace}, and OpenSea~\cite{OpenSea_Marketplace}. With OpenSea that allows trading in multiple cryptocurrencies. We also monitored the WAX blockchain, through tracking transactions in the Atomic API~\cite{Atomic_Marketplace}. 

We group NFTs into six categories: \emph{Art} consisting of digital artworks such as images, videos, or GIFs; \emph{Collectible} representing items of interest to collectors; \emph{Games} including digital object used in competitive games; \emph{Metaverse} consisting of pieces of virtual worlds; \emph{Utility} representing items having a specific function; and \emph{Other} including the remaining collections. More details on the NFT categorization are explained in Section~\ref{data_cleaning_categorization}. The final, cleaned dataset includes 935 million USD traded in 6.1 million transactions involving 4.7 million NFTs grouped in 4,624 collections. Our dataset includes transactions in 160 different cryptocurrencies with most of them made in WAX (52\% of the total number of transactions), while the volume in USD is mostly ETH (81\% of the total volume). Table~\ref{Statistics_NFTs} shows general statistics of the categories of NFTs considered, involving a total of 359,561 buyers, 314,439 sellers, trading 4.7 millions NFTs involving 953 million USD in cryptocurrencies.

\subsubsection{Image collection and visual feature extraction}
For each NFT in our dataset (except for less than 3\,000 exceptions) we managed to collect at least one URL that points to a copy of the NFT's digital object. We focused only on objects with image file formats (e.g. PNG, SVG, JPEG) and GIFs, for a total of about 1.2 million unique graphical objects associated with 4.7 million unique NFTs. Note that a single digital object can be related to multiple NFTs; this happens for example for identical playing cards that are minted in multiple copies, each associated with a different NFT. Since our algorithm for visual feature extraction works with static images, we converted the animated GIFs to PNGs by extracting central frame of each GIF.
In order to succinctly represent the visual features that characterize an image, we encode it into a latent space using a neural network. Specifically, we pick the PyTorch~\cite{paszke2019pytorch} implementation of AlexNet~\cite{krizhevsky2012imagenet}, a deep convolutional neural network architecture designed for image classification. We initialize AlexNet with weights pre-trained on ImageNet~\cite{deng2009imagenet}, a widely-used reference dataset of labeled images. Given an image in input, AlexNet passes it through multiple layers of transformation. The second to last layer (i.e., the layer before the classification layer) is a vector consisting of $4\,096$ values that constitute a dense representation of the input image into a high-dimensional space. These vectors can be used for a variety of tasks such as similarity ranking, clustering, or classification. To reduce the dimensionality of AlexNet vectors, we extracted their principal components using Principal Component Analysis (PCA)~\cite{jolliffe1986principal}, and selected the 5 most relevant ones. PCA projects each point of the high-dimensional space into a space with a desired number of dimensions, while preserving the data variation as much as possible. 

\section*{Acknowledgements}
The authors are grateful to NonFungible Corporation for helpful conversations and data sharing (see text). The research was partly supported by The Alan Turing Institute.

\section*{Data availability}

Data and scripts used to download from the different APIs will be available upon publication.

\bibliographystyle{unsrt}

\clearpage

\setcounter{table}{0}
\setcounter{figure}{0}
\setcounter{section}{0}

\renewcommand{\thesection}{S\arabic{section}}  
\renewcommand{\thetable}{S\arabic{table}}  
\renewcommand{\thefigure}{S\arabic{figure}}

\begin{center}
\large{\textbf{Supplementary Information}}
\end{center}
\section{Additional Analyses}
\label{Additional_Analyses}

\begin{figure}[h]
\includegraphics[width=\textwidth]{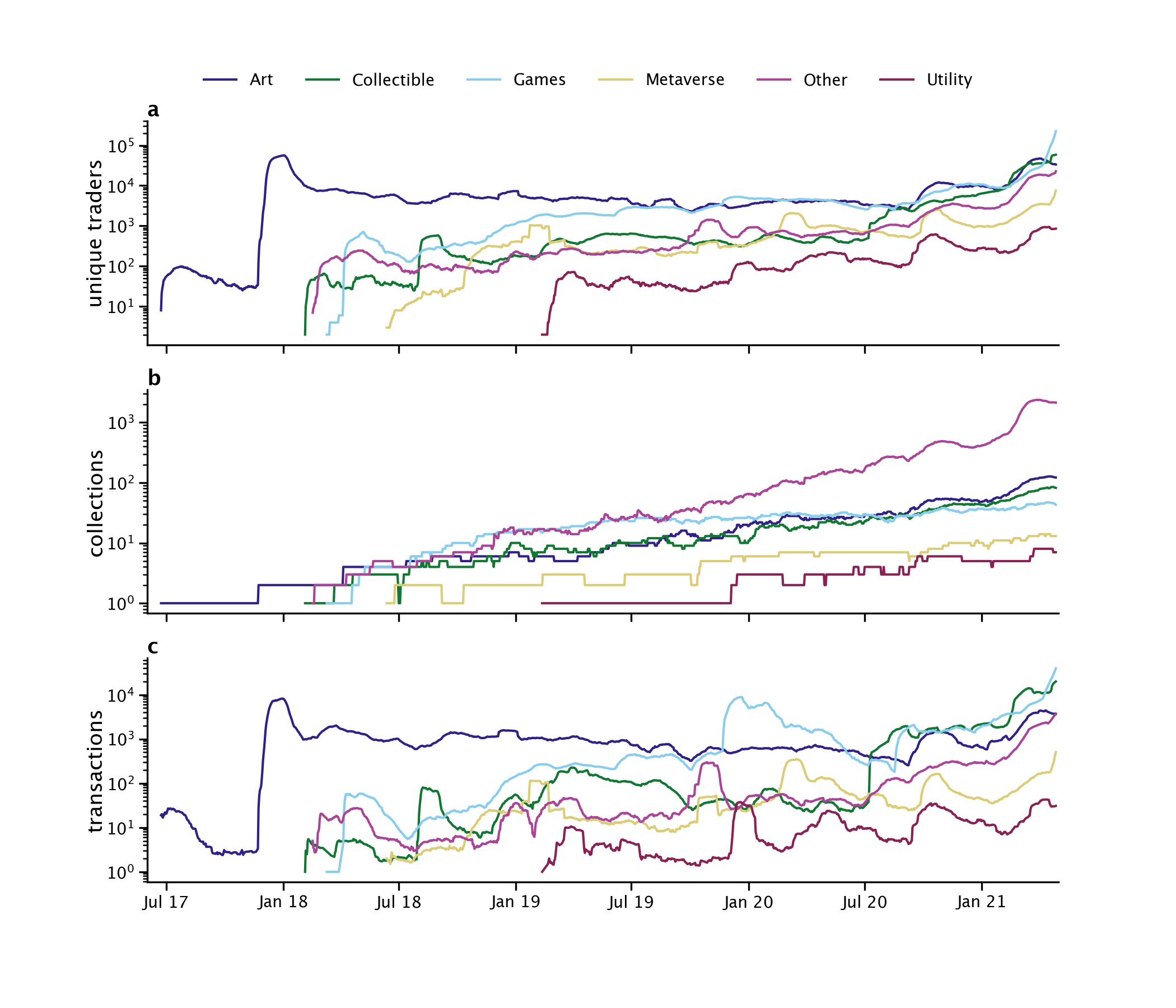}
\caption{\textbf{Evolution of the NFT market.} The number of unique traders \textbf{(a)}, collections \textbf{(b)} and transactions \textbf{(c)} over time for different categories. Results are computed over a rolling window of 30 days.}
\label{FigS1}
\end{figure}

\begin{figure}[h]
\includegraphics[width=\textwidth]{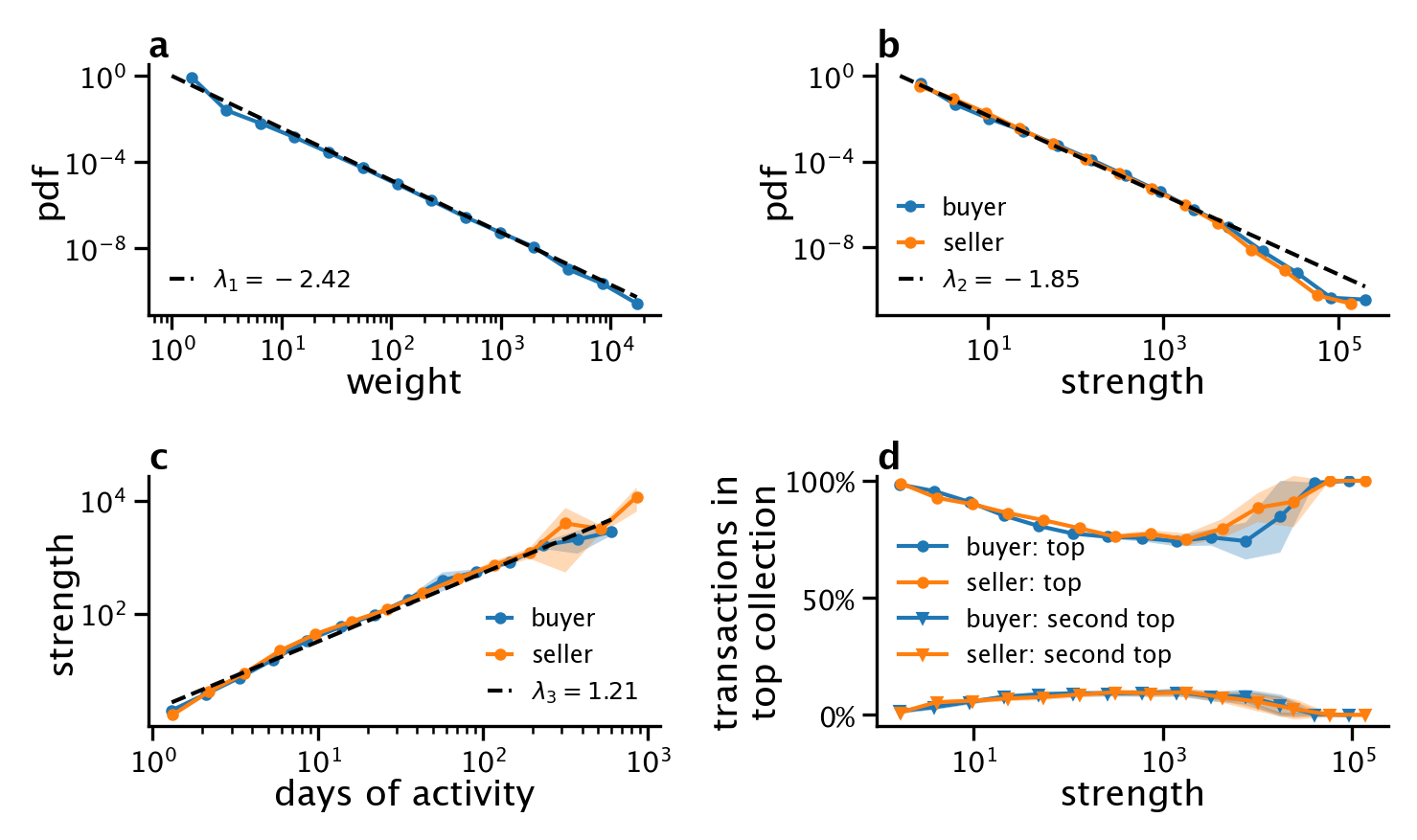}
\caption{\textbf{Key network properties of buyers and sellers.} \textbf{(a)} Probability distribution function of the number of transactions (weight) from buyers to sellers. \textbf{(b)} Probability distribution function of the buyers and sellers' strength. \textbf{(c)} Relationship between the buyers and sellers' strength and the number of days in which they are active. \textbf{(d)} Percentage of transaction buyers and sellers make toward their top and second-top NFT collections.}
\label{FigS2_network}
\end{figure}

\begin{figure}[h]
\includegraphics[width=\textwidth]{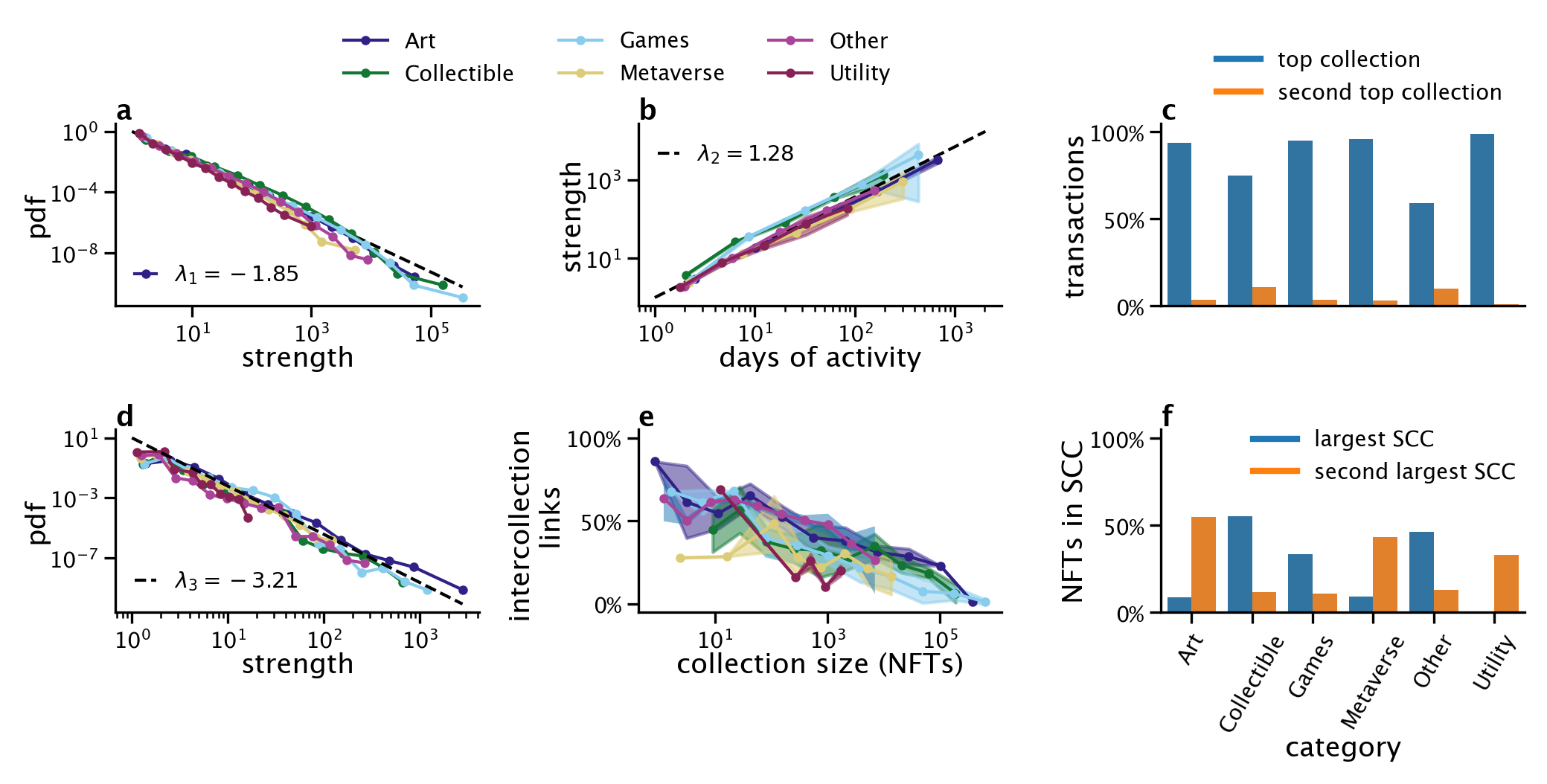}
\caption{\textbf{Key network properties for each category.} \textbf{(a)} Probability distribution function of the traders' strength. \textbf{(b)} Relationship between the traders' strength and the number of days in which they are active. \textbf{(c)} Percentage of transactions all traders make toward their top and second-top NFT collections. \textbf{(d)} Probability distribution function of the NFTs' strength. \textbf{(e)} Percentage of transactions between NFTs in different collections as a function of the size of the collection. \textbf{(f)} Percentage of NFTs belonging to the first and second largest strong connected component (SCC). Solid curves in panels \textbf{(b)-(e)} represent average values, while respective bands the 95\% confidence interval.}
\label{Fig_networks_categories}
\end{figure}

\begin{figure}[h]
\includegraphics[width=\textwidth]{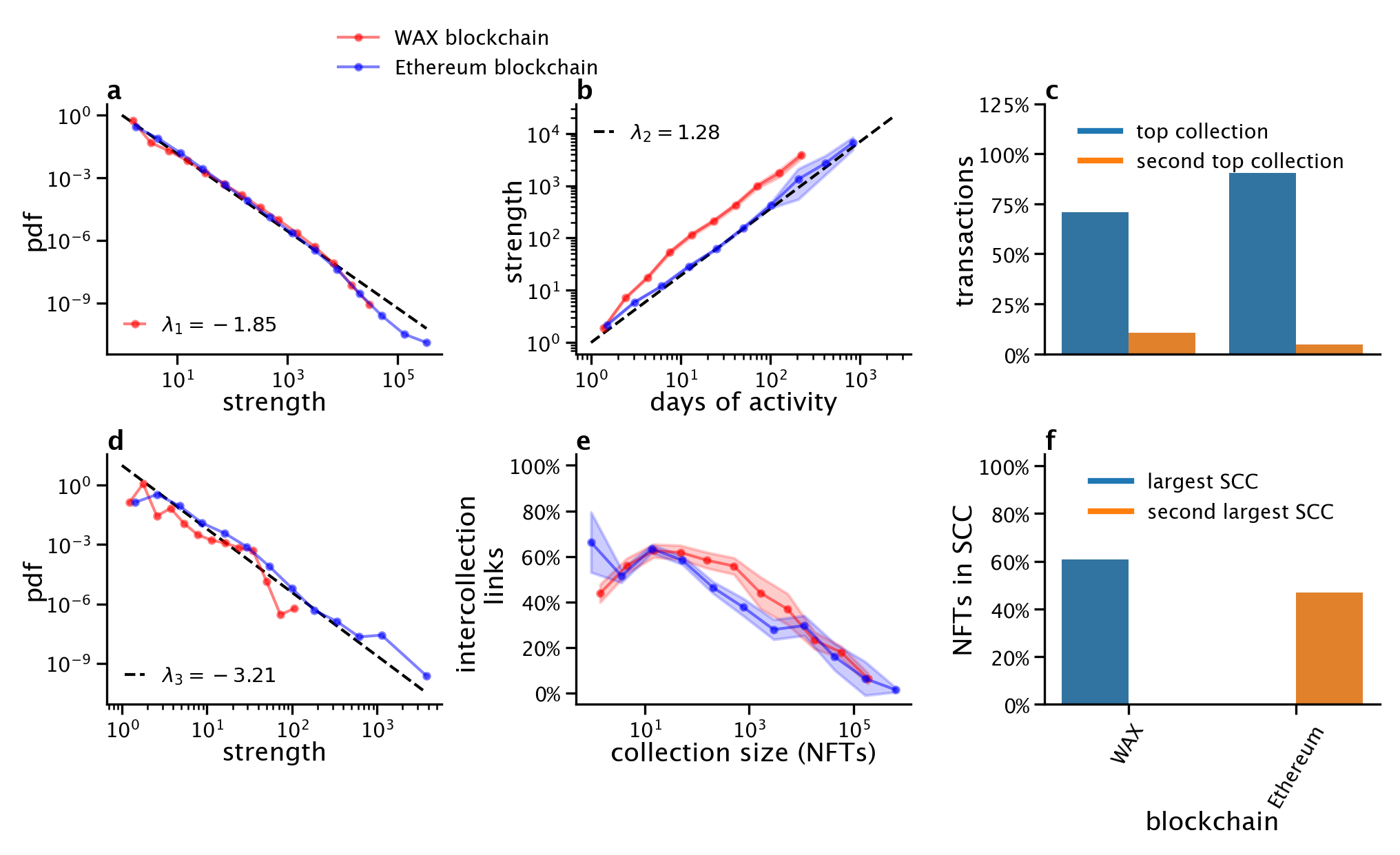}
\caption{\textbf{Key network properties for each blockchain.} \textbf{(a)} Probability distribution function of the traders' strength. \textbf{(b)} Relationship between the traders' strength and the number of days in which they are active. \textbf{(c)} Percentage of transactions all traders make toward their top and second-top NFT collections. \textbf{(d)} Probability distribution function of the NFTs' strength. \textbf{(e)} Percentage of transactions between NFTs in different collections as a function of the size of the collection. \textbf{(f)} Percentage of NFTs belonging to the first and second largest strong connected component (SCC). Solid curves in panels \textbf{(b)-(e)} represent average values, while respective bands the 95\% confidence interval.}
\label{Fig_networks_blockchain}
\end{figure}

\begin{figure}[h]
\includegraphics[width=\textwidth]{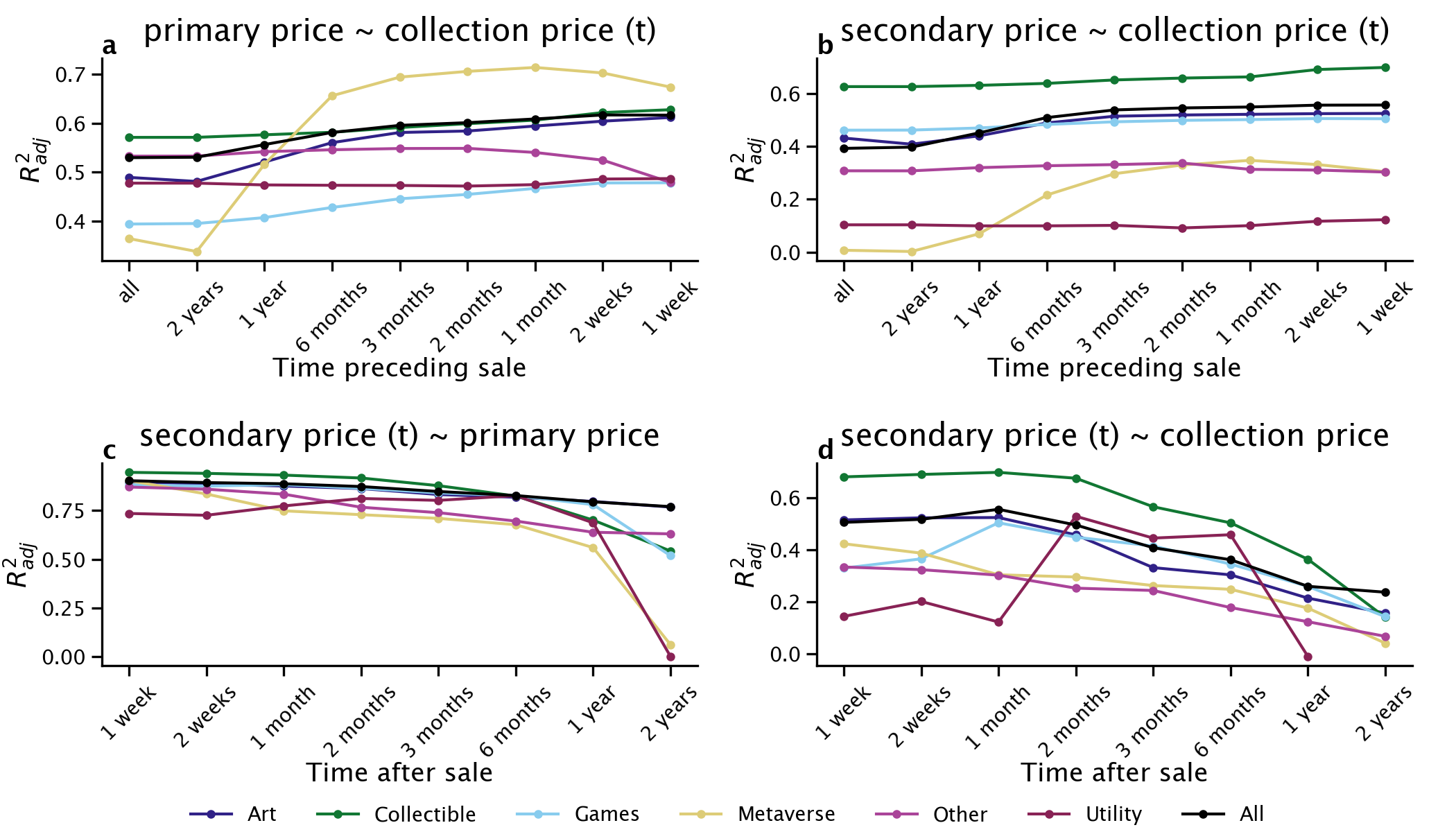}
\caption{\textbf{Primary and secondary sale price predictions.} Top: $R_{adj}^2$ of a linear regression fit to predict \textbf{(a)} the price of primary sales, and \textbf{(b)} the median price of secondary sales 1 month after their respective primary sale from the historical median price of sale in the collection calculated over varying time windows (one week to two years) preceding the primary sale. Bottom: $R_{adj}^2$ of a linear regression fit to predict \textbf{(c)} the price of secondary sales from the price of their respective primary sales, and \textbf{(d)} the price of secondary sales from the median price of sales in the NFT's collection in the previous week; we perform different regressions to predict the median price of secondary sales over varying time windows (one week to two years) after the primary sale. All results are broken down by NFT categories.}
\label{fig:regression_vs_window_before}
\end{figure}

\begin{figure}[h]
\includegraphics[width=0.5\textwidth]{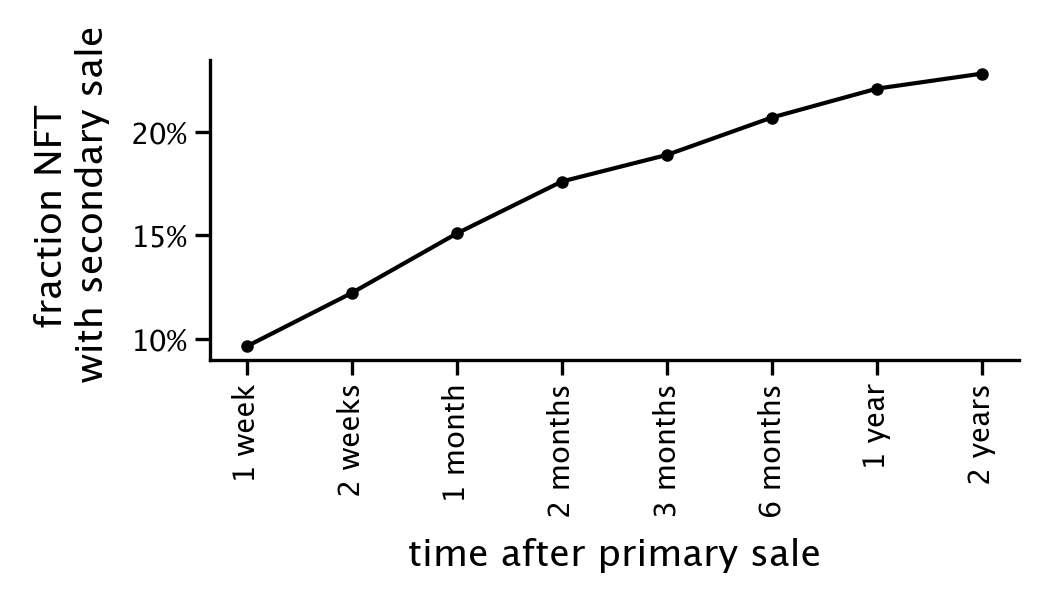}
\caption{\textbf{Vast majority of NFTs does not have a secondary sale.} Fraction of NFTs that were sold in at least one secondary sale $n$ days after their primary sale.}
\label{fig:fraction_nft_with_resale}
\end{figure}

\begin{figure}[h]
\includegraphics[width=0.5\textwidth]{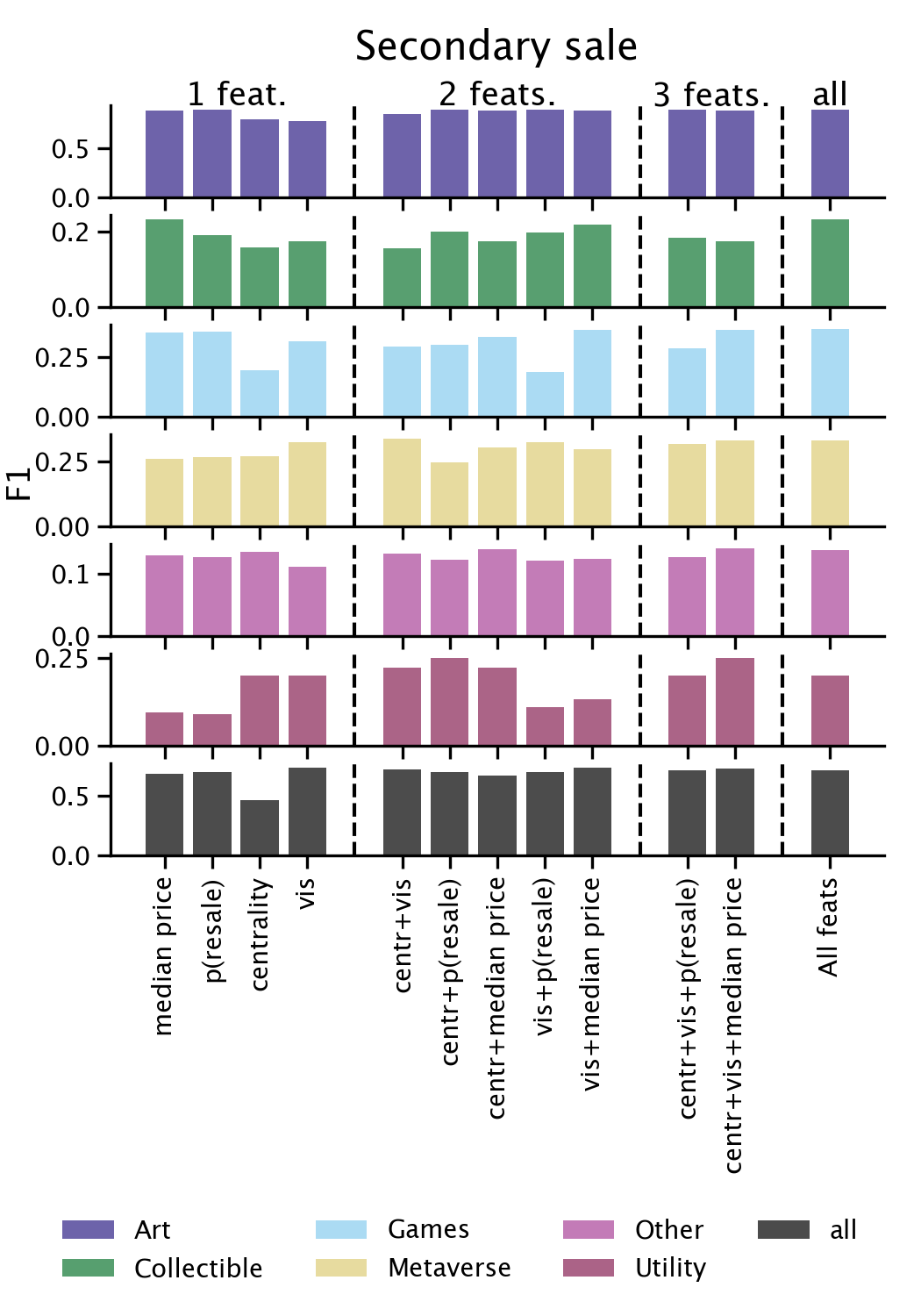}\\
\caption{\textbf{Result of predicting the existence of a secondary sale.} $F1$ score of a binary classification task aimed at predicting whether a NFT will be sold in a secondary sale within 1 year after its primary sale. Results are broken down by different feature sets and NFT categories.}
\label{fig:prediction_features_breakdown}
\end{figure}

\begin{figure}[h]
\includegraphics[width=0.5\textwidth]{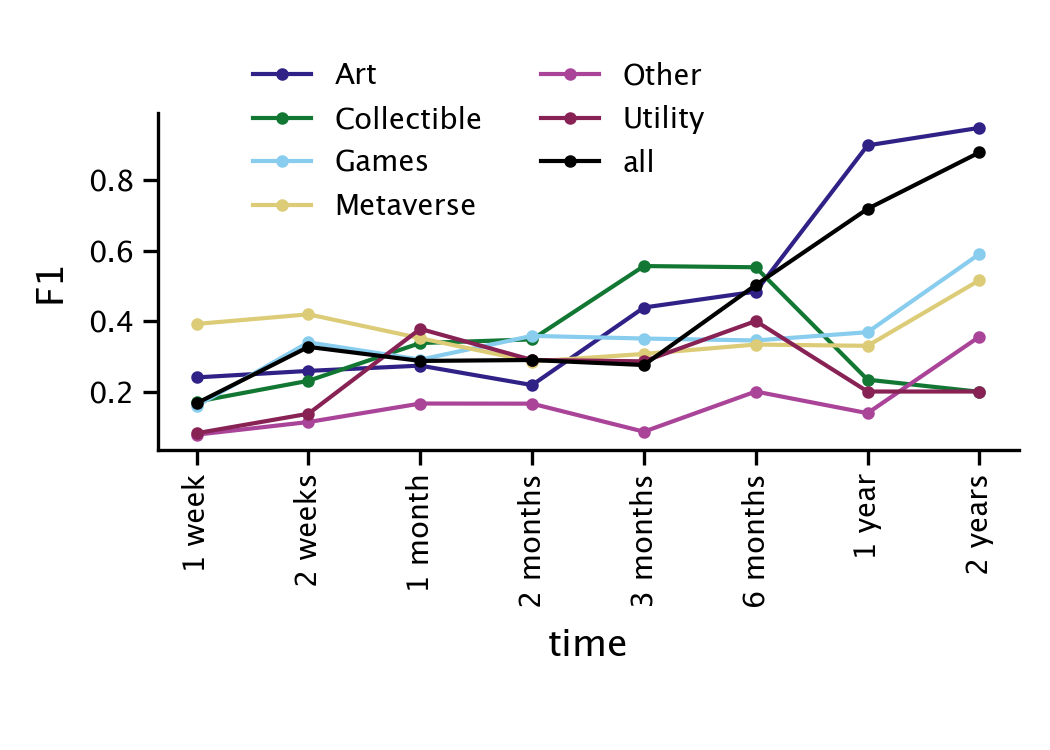}\\
\caption{\textbf{Classification task with different time windows.} $F1$ score of a binary classification task aimed at predicting whether a NFT will be sold in a secondary sale within varying time windows after its primary sale. We used all available features for training and testing the models. Results are broken down by different NFT categories.}
\label{fig:prediction_allfeats_vs_time}
\end{figure}

\clearpage

\section{Additional information on the data and Methods}
\label{Data_methods_additional}

\begin{table}
\footnotesize
\setlength{\tabcolsep}{5pt}
\begin{tabular}{lcccc}
\textbf{Category} &  \textbf{Buyers} &  \textbf{Sellers} &   \textbf{NFTs} &  \textbf{Volume ($\cdot 10^{6}$ USD)} \\
\hline
\hline
\emph{Art}         &  161,423 &    70,623 &   859,570 &            655.62 \\
\emph{Collectible} &   62,100 &    67,173 &  1,344,449 &            109.84 \\
\emph{Games}       &  151,702 &   192,772 &  2,202,432 &            70.77 \\
\emph{Metaverse}   &   12,121 &    10,283 &    47,286 &            68.18 \\
\emph{Utility}     &    2,637 &     1,483 &     7,752 &           8.74 \\
\emph{Other}       &   34,647 &    22,308 &   242,990 &            21.96 \\
\hline
Total & 359,561 & 314,439 & 4,704,479 & 935.11 \\
\end{tabular}
\caption{\textbf{Breakdown of NFTs categories.} Overall statistics of each NFT category under consideration.}
\label{Statistics_NFTs}
\end{table}

\subsection{Data cleaning and categorization}
\label{data_cleaning_categorization}

NFTs that share common features are grouped in \emph{collections}, which names are cleaned and even out. The raw names, as downloaded from the selected sources, are stripped by any digits, special characters (e.g., ``-''), unusual patters (e.g., ``xxxxx''), and capitalized. Cleaned names are then even out by considering a \href{https://drive.google.com/file/d/1mbHUKdonNPpesvwGeXZHSl8qKkr-u4LS/view?usp=sharing}{list of words}. For instance, the collection Aavegotchi renames all collections starting with that string of characters in Aavegotchi. Some other collections with generic names (e.g. Stuff) are called Miscellanea. 

Fields considered in our analysis are: buyer address, seller address, time of the transaction, name of the collection, ID of an NFT (here simply called ``NFT''), url to the NFT's digital object, type of cryptocurrency and its amount used in the transaction. Transactions with one of the former fields empty (except for the url to the NFT's digital object) are removed from the dataset. From these remaining data, the price in USD is computed considering the exchange rate of the given cryptocurrency at the day of the transaction. Note that, in this work, we use buyer or seller addresses as proxies for real identities, as commonly done in the Ethereum blockchain~\cite{Cryptopunks_leaderboard} and with the usernames~\cite{NFT_NYT_Barabasi}, while in reality an individual may have multiple addresses or usernames. 
NFTs sharing common features, such as, digital cards of the same online game, belong to the same collection. Also, collections are assigned to one of the following six categories: \emph{Art}, \emph{Collectible}, \emph{Games}, \emph{Metaverse}, \emph{Utility}, or \emph{Other}. The operative definitions of these categories are inspired from the definitions given by NonFungible Corporation~\cite{nonfungible_data}, a specialized company that track NFTs sales, and OpenSea~\cite{OpenSea_assets}, one of the largest NFT marketplace, and summarized in Table~\ref{categorization_NFTs}. 

All collections with high trading volume or large number of sales were categorized by at least two authors of the present manuscript. The manual categorization was done independently by each author, then the final category selected by majority voting, asking the opinion of additional authors in case of draw between two or more categories. Note that a collection may belong to more than one category and forcing each collection into one category only is a limitation of the present work.

With the exception of the Atomic API, the downloaded datasets are not independent and, for instance, some transactions shared by NonFungible Corporation are available from OpenSea as well. When data are merged together, duplicated transactions are removed by prioritizing (in order) data from NonFungible Corporation, CryptoKitties sales, Gods-Unchained API, Decentraland API, and OpenSea API.

\begin{table}[!t]
\footnotesize
\setlength{\tabcolsep}{5pt}
      \begin{tabular}{l|c}
        \hline
        \hline
          \textbf{Category} & \textbf{Description} \\
           \hline
         \hline
           \thead{\emph{Art}} & \thead{NFTs of digital artworks, such as images, videos, or GIFs}   \\
           \thead{\emph{Collectible}} & \thead{NFTs of interest to a collector}  \\
           \thead{\emph{Games}} & \thead{NFTs used n competitive games}  \\
           \thead{\emph{Utility}} & \thead{NFTs for specific purposes (e.g. secure and decentralized name service)}  \\
           \thead{\emph{Metaverse}} & \thead{Piece of virtual worlds} \\
           \thead{\emph{Other}} & \thead{NFTs of small collections that are not included in the other categories}  \\
      \end{tabular}
\caption{\textbf{NFTs categorization.} Operative definitions of NFTs categories.}
\label{categorization_NFTs}
\end{table}

\subsection{Generation of the traders and NFTs networks of interaction}
\label{Trader_NFT_networks}

\begin{table}[h]
\footnotesize
\setlength{\tabcolsep}{5pt}
\begin{tabular}{cccccc}
\textbf{Example} & \multicolumn{3}{c}{\textbf{Trader network}} & \multicolumn{2}{c}{\textbf{NFT network}} \\
\hline
\hline
(i) & 
\subfloat[$t=t_\alpha$]{ \includegraphics[width=0.13\textwidth]{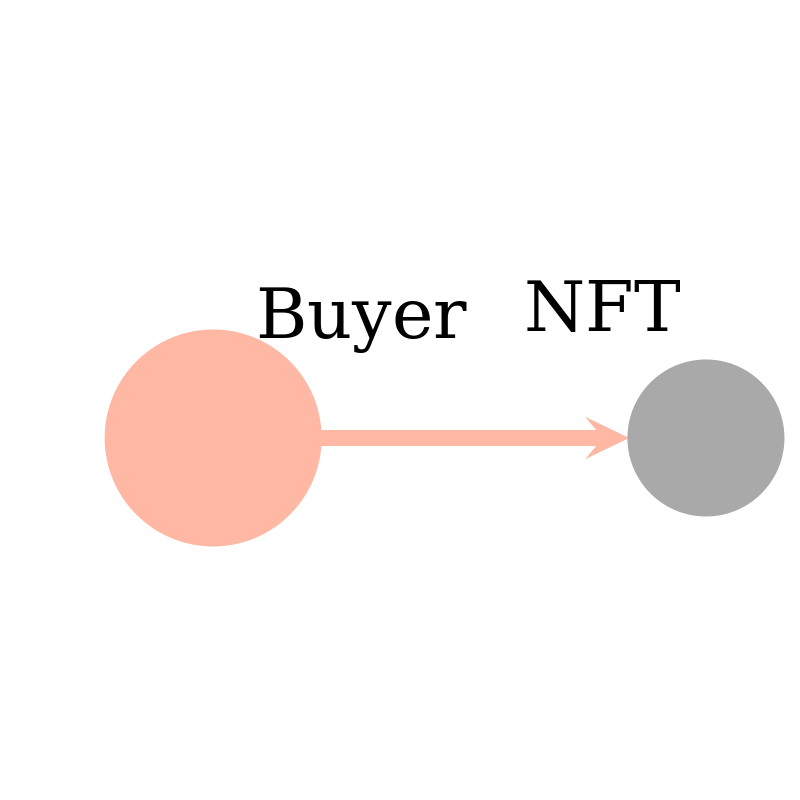}} &
\subfloat[$t=t_\beta>t_\alpha$]{ \includegraphics[width=0.13\textwidth]{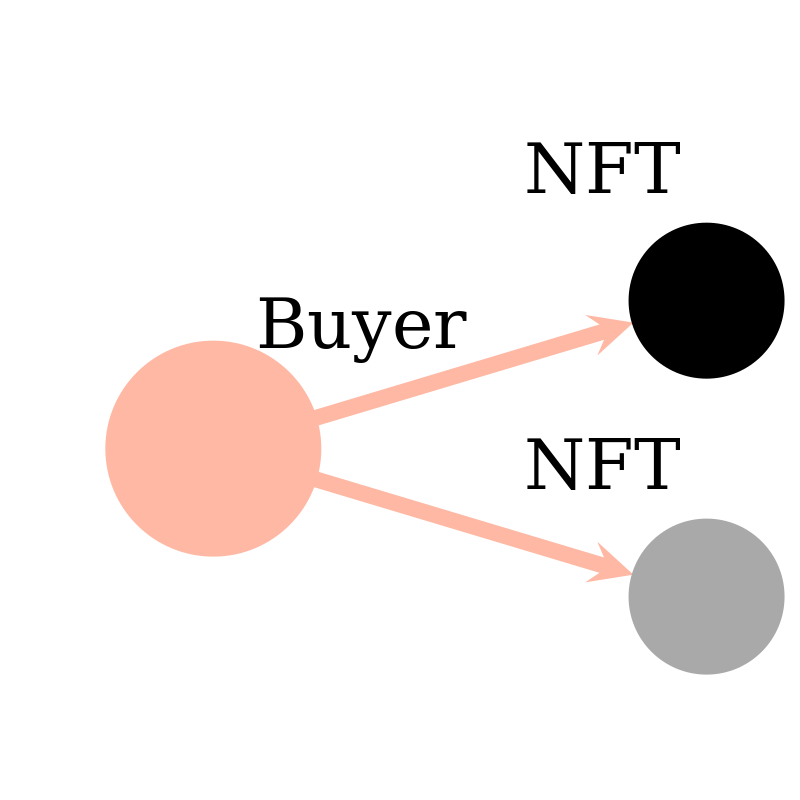}} & \subfloat[$t=t_\gamma>t_\beta$]{ \includegraphics[width=0.13\textwidth]{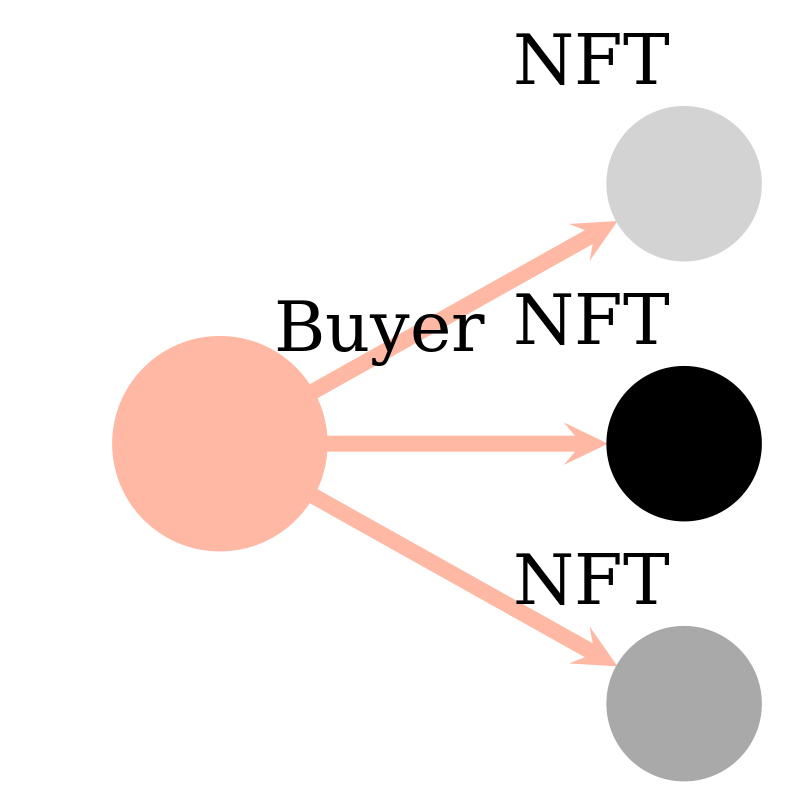}} &
\subfloat[$t=t_\beta$]{ \includegraphics[width=0.13\textwidth]{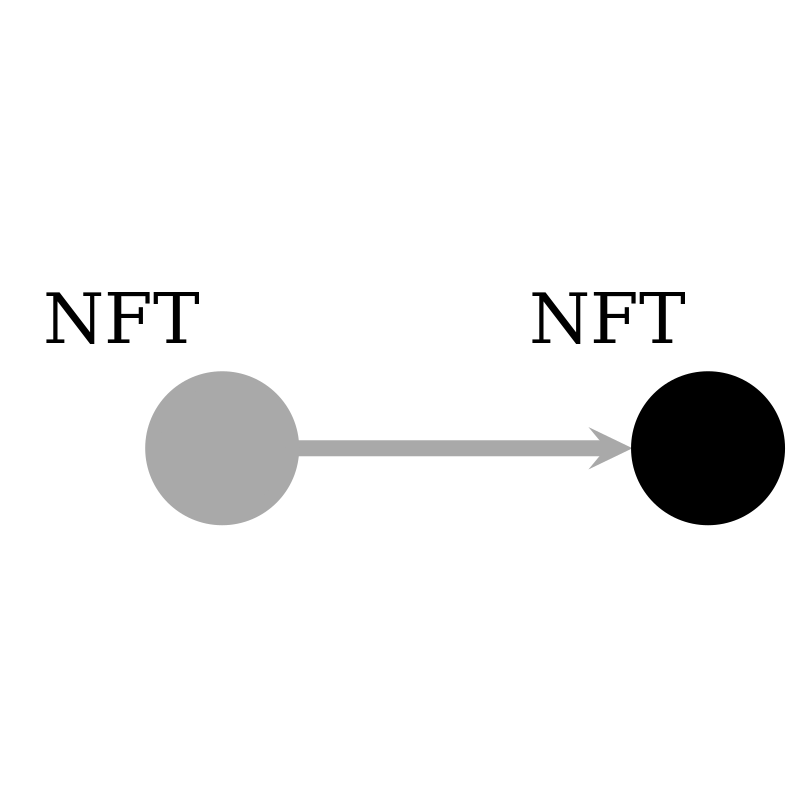}} & \subfloat[$t=t_\beta^{\prime}$]{ \includegraphics[width=0.13\textwidth]{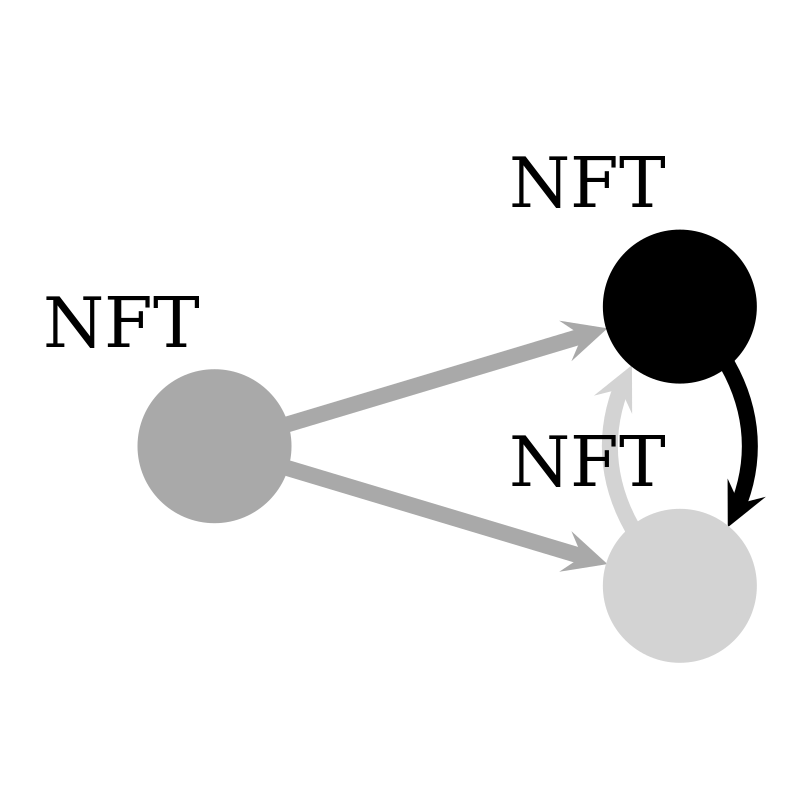}} \\
(ii) & \subfloat[$t=t_\alpha^{\prime}$]{ \includegraphics[width=0.13\textwidth]{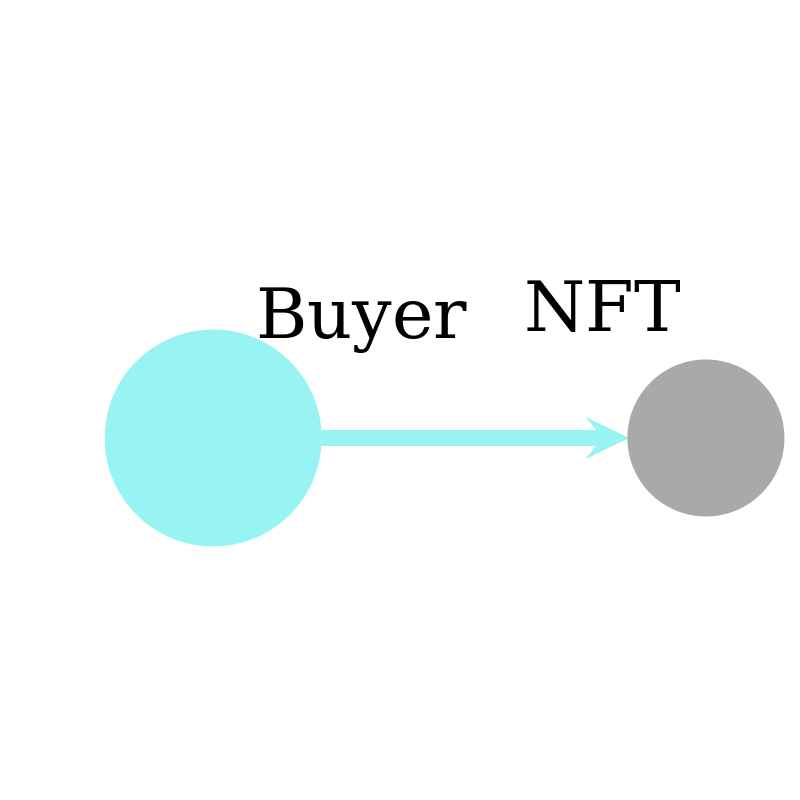}} & \subfloat[$t=t_\beta^{\prime}>t_\alpha^{\prime}$]{\includegraphics[width=0.13\textwidth]{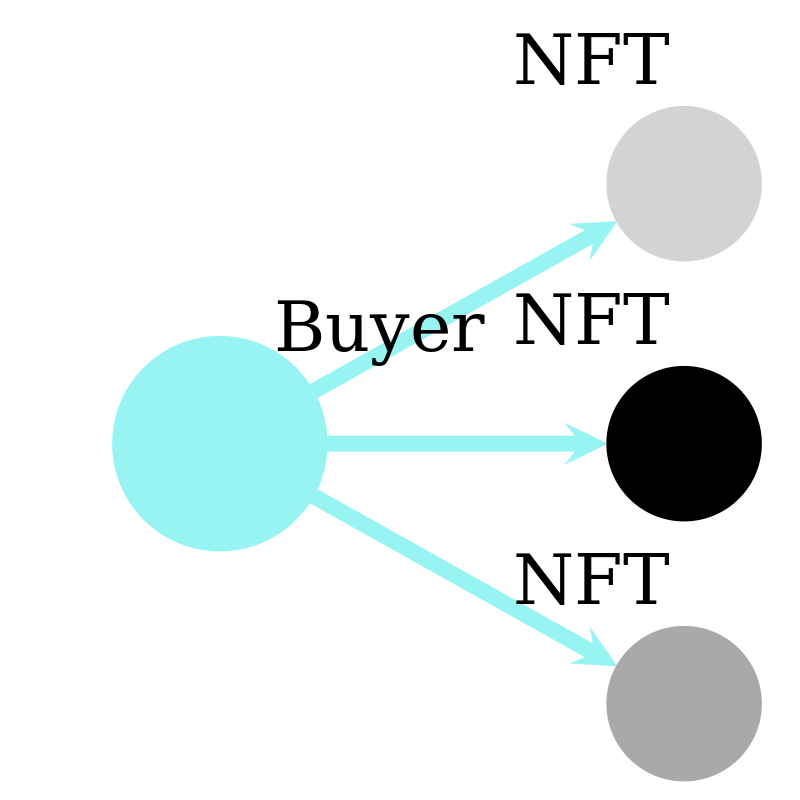}} & \subfloat[$t=t_\gamma^{\prime}>t_\beta^{\prime}$]{ \includegraphics[width=0.13\textwidth]{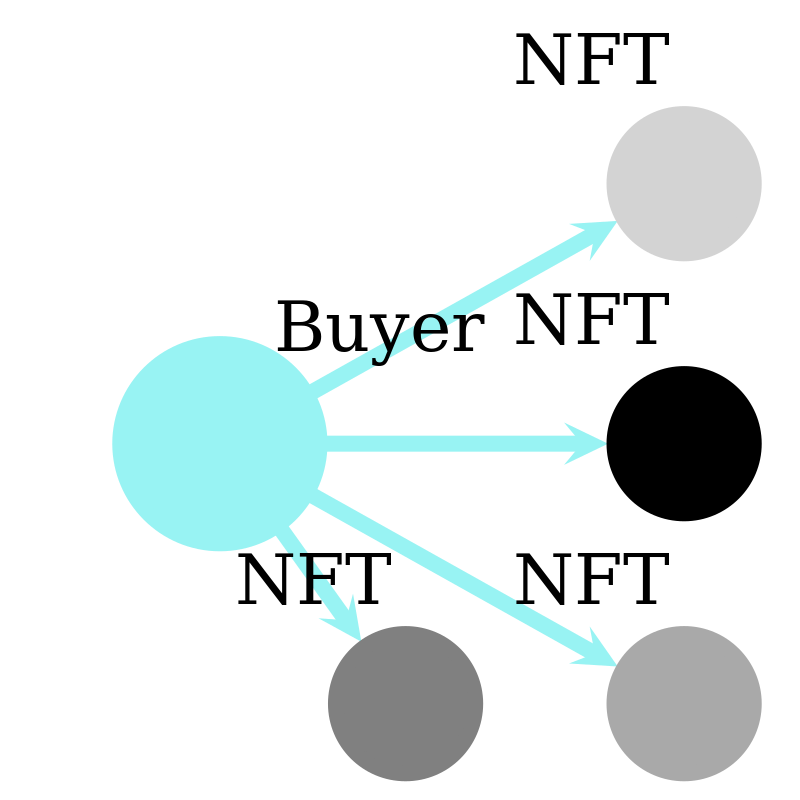}}  & \subfloat[$t=t_\beta^{\prime}$]{ \includegraphics[width=0.13\textwidth]{NFT_multiple_links.png}} & \subfloat[$t=t_\gamma^{\prime}$]{ \includegraphics[width=0.13\textwidth]{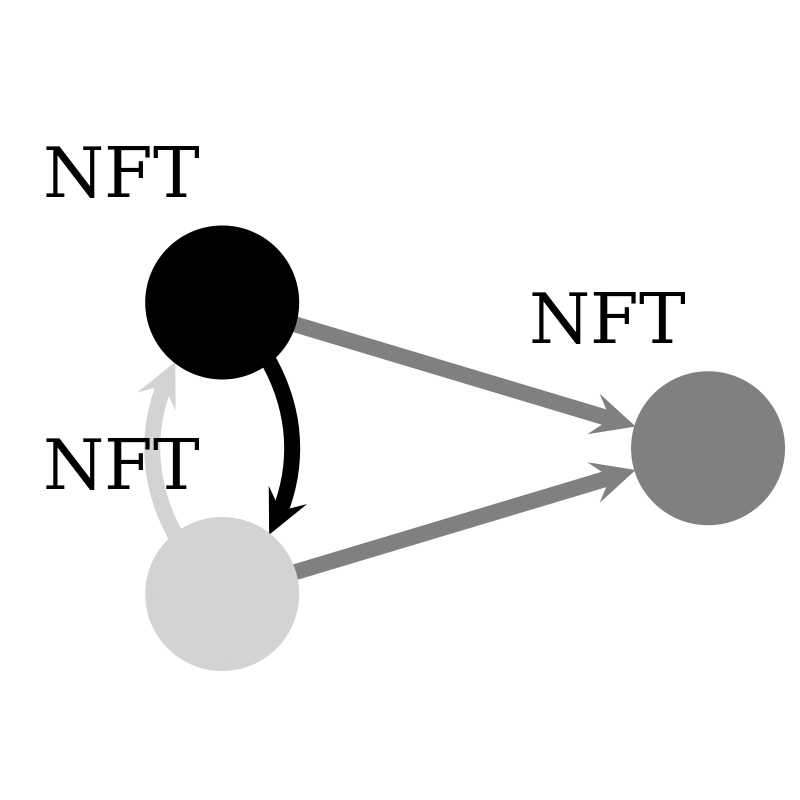}}\\
\end{tabular}
\caption{\textbf{Link creation mechanism of the NFT network.} Directed links are generated using the trader network as reference and following three rules. The first two rules take into consideration the same buyer, while the third rule another buyer, both interacting with the same three NFTs. Visualization is done using Graph-tool:~\cite{peixoto_graph-tool_2014}.}
\label{Rules_NFT_network}
\end{table}

While the trader network was directly obtained from our data collection, the NFT network was created by linking NFTs that are purchased in a sequential order by the same buyer. Let's consider NFT$_i$, NFT$_j$, NFT$_k$, and NFT$_h$ as identifier of generic NFTs, and $t_\alpha$, $t_\beta$, $t_\gamma$, $t_\alpha^{\prime}$, $t_\beta^{\prime}$, and $t_\gamma^{\prime}$ as identifiers for time instants (with a temporal resolution of seconds). Table~\ref{Rules_NFT_network} illustrates two meaningful examples of how the NFT network is created. (i) When a buyer, who purchased NFT$_i$ at time $t_\alpha$, buy NFT$_j$ at time $t_\beta>t_\alpha$, a directed link from NFT$_i$ to NFT$_j$ is created at time $t_\beta$. If the same buyer purchases NFT$_k$ at a later time $t_\gamma>t_\beta$, a directed link from NFT$_j$ to NFT$_k$ is drawn at time $t_\gamma$. (ii) When a buyer, who purchased NFT$_i$ at time $t_\alpha^{\prime}$ buy NFT$_j$ and NFT$_k$ at the same instant $t_\beta^{\prime}>t_\alpha^{\prime}$, a directed link from NFT$_i$ to NFT$_j$ and another from NFT$_i$ to NFT$_k$ are drawn. If the same buyer purchases a fourth NFT$_h$ at time $t_\alpha^{\prime}$. The NFT network hereby constructed includes $4\,657\,713$ NFTs out of a total of $4\,704\,479$. The NFTs that are left out belongs to buyers who perform only one transaction. The network analysis is done by leveraging selected functions in the networkx Python package. 

\subsection{Generation of random networks}
\label{sec:random_network}

Random networks relative to the trader and NFT networks are generated in a similar fashion, and by preserving each node outgoing and incoming strength. We consider the pool of observed links with repetition, that is, a link appears a number of times equal to its weight. Two links ($l_1$ and $l_2$) are randomly extracted over this pool, where node $n_{l_1}^1$ create a directed link to node $n_{l_1}^2$ and node $n_{l_2}^1$ create a directed link to node $n_{l_2}^2$. These links are swapped if the fours nodes are different. The swap consists in creating link $l_1^{\prime}$, where node $n_{l_1}^1$ create a directed link to node $n_{l_2}^2$, and link $l_2^{\prime}$, where node $n_{l_2}^1$ create a directed link to node $n_{l_1}^2$. We repeat the procedure for a number of times equal to the total links in the network. We create 100 independent realization of this random network for the trader network and 100 for the NFT network.

\subsection{NFT features}
\label{sec:features}

We characterize NFTs with a set of 11 features, partitioned in three groups. An NFT's features were calculated only from the data that could be collected until the day before its primary sale, $t_s$. We used these features in two separate tasks of regression (Section~\ref{sec:regression}), and classification (Section~\ref{sec:classification}).

The first group of features includes network centrality scores obtained from the trader network. Specifically, we considered the degree centrality ($k$), and the PageRank centrality ($PR$) of the seller and the buyer, for a total of 4 features. The degree centrality of a node is the count of all its incoming and outgoing unique links~\cite{wasserman1994social}, and its PageRank centrality measures the stationary probability that a random walk on the network ends up in that node~\cite{brin1998anatomy}.

The second group includes the visual features of the object associated with the NFT, namely 5 PCA components extracted from the AlexNet vector of the object ($PCA_{1\ldots5}$). We experimented with a number of components varying from 2 to 10, and results varied only slightly---fewer components caused a feeble decrease in the quality of the regression and prediction results, while additional components did not add any predictive power.

The third and last group includes two features to account for the previous sale history in the NFT's collection. The first is the median price of primary and secondary sales made in the collection of interest during a time window prior to $t_s$. The latter models the prior probability of secondary sale. We acknowledge that the likelihood that a NFT gets transacted in a secondary sale might depend on the collection it belongs to. For example, NTFs corresponding to collectible items from very popular collections may be more likely to be resold than an NFT serving for a specific purpose, such as determining the ownership of a name server. We defined the probability of secondary sale, $p_{resale}$, as $0.5$ (random probability) when the NFT is the first to be sold in its collection; else, the probability of secondary sale is calculated as:
\begin{equation*}
p_{resale} = \frac{0.5}{n+1}+\frac{n}{n+1}\frac{s}{n},
\label{eqn:p_resale}
\end{equation*}
where $n$ represents the NFTs with a primary sale up to the day before the first purchase and $s$ the number of these NFTs with at least one secondary sale. When the collection is large, the probability of secondary sales becomes $p \left(n \rightarrow + \infty \right) = s/n$ and corresponds to the ratio between items with secondary sales over all items with one sale.

The frequency distributions of our features have different skews and ranges. To make them comparable and suitable for regression and prediction tasks, we first transform their values to make their distributions closer to a Normal distribution. Specifically, we calculate the logarithm of the network degree and the median sale price (after adding 1, so that zero-values were preserved), and we apply a BoxCox transformation~\cite{box1964analysis} to the PageRank centrality and to $p_{resale}$; BoxCox uses power functions to create a monotonic transformation that stabilizes variance and makes the data closer to a normal distribution. No transformation was needed for the PCA features. Last, we scale all the variables in the range $[0,1]$ (i.e., min-max scaling).

\subsection{Sale price regression}
\label{sec:regression}

We perform linear regressions to estimate an NFT's primary and secondary sale prices. Linear regression is an approach for modeling a linear relationship between a dependent variable (secondary sale price, in our experiments) and a set of independent variables (features describing the NFT at the time it was first sold), and it does so by associating a so-called $\beta$-coefficient with each independent variable such as the sum of all independent variables multiplied by their respective $\beta$-coefficients approximates the value of the dependent variable with minimal error. Specifically, we used an Ordinary Least Squares regression model to estimate coefficients such that the sum of the squared residuals between the estimation and the actual value is minimized.

We use the NFT features described in Section~\ref{sec:features} as independent variables, and either the price of primary sale or the median secondary sale price calculated over a time window starting at $t_s$ as dependent variables. For secondary sale price, the results changed only slightly when using different aggregations other than the median (e.g., mean, maximum). We experimented with different lengths of the time window, ranging from one week after the primary sale up to two years after. To make sure that the secondary sale price of each NFT was calculated over time windows of equal length, we exlcuded from the regression NFTs that were sold for the first time too recently---namely those NFTs whose $t_s$ was within one time window before the most recent timestamp in our dataset. In the regression, we considered only NFTs with at least one secondary sale in the time window considered.

We evaluated the goodness of the linear fit using coefficient of determination $R^2$, a score in the range $[0,1]$ that measures the proportion of the variance in the dependent variable that the linear model is able to predict from the independent variables. In particular, we used its `adjusted' version $R_{adj}^2$, that discounts the effect of the $R^2$ spuriously increasing as more independent variables are added to the model.

\subsection{Secondary sale prediction}
\label{sec:classification}

We performed a binary classification task to predict whether an NFT will be transacted in a secondary sale after its primary sale at time $t_s$. We adopted a standard supervised learning approach. In supervised learning, instances in a dataset (the NFTs) are described with a number of features (those presented in Section~\ref{sec:features}) and marked with a target label (1 if the NFT was transacted in a secondary sale, 0 otherwise). A mathematical model learns a function that maps the features to the target label based on a number of \emph{training} instances from the dataset. The performance of the model is later assessed on a \emph{test} set of unseen instances. In our experiments, we emulate a prediction on future data based on past knowledge. To do so, we sort the NFTs according to their time of primary sale $t_s$, and we use the first 95\% of NFTs for training and the latest $5\%$ for testing. Our dataset is sufficiently large so that the test set, albeit small in relative terms, includes a large selection of tens of thousands of instances. Similar to the regression task, we consider multiple time windows of varying size to determine the target label (i.e., whether the NFT was resold or not), and we exclude from the dataset recent NFTs whose $t_s$ is within one time window before the last timestamp in our dataset.

There are several classes of models that can be used for supervised learning~\cite{hastie2009elements}. We pick AdaBoost~\cite{freund1999short}, an ensemble of weak learners (in our case, decision trees) whose output is combined into single score through a weighted sum. In particular, we initialized the AdaBoost classifier with 100 decision tree stumps (i.e., trees of depth 1), and trained it with a learning rate of 1. Despite its relatively simple design, AdaBoost can achieve good performance compared to more complex model and it effectively limits overfitting the learned function on the training data.

The labels of our dataset are \emph{imbalanced}: the number of negative labels is much higher than the number of positive ones (i.e., 80\% of NFTs in our dataset are more not resold). Imbalanced datasets can affect the ability of the model to learn a function that can effectively associate the correct label to both positive and negative instances. To mitigate this problem, we perform random oversampling~\cite{ling1998data} to balance the classes. Specifically, within the training set, we add multiple copies of positive samples picked at random until the size of the two classes is balanced. Compared to other oversampling techniques~\cite{chawla2002smote,he2008adasyn}, random oversampling does not generate synthetic data points, which exhibiting unrealistic features. By applying oversampling, we effectively set the model to assign higher importance to positive samples: misclassifying a positive instance causes a loss in performance that is proportional to the number of its replicas.

To evaluate the performance on the test set, we measure two quantities. The first is the \emph{F1-score}, namely the harmonic mean of the precision (fraction of instances that are classified as positive that are indeed positive) and recall (fraction of positive instances the are correctly classified). The second is the ``Area Under the ROC Curve'' (AUC); it measures the ability of the model to correctly rank positive and negative samples by confidence score, independent of any fixed decision threshold. AUC is equal to 0.5 for a random classification and it is equal to 1 for a perfect ranking.

\end{document}